\documentclass[10pt,twocolumn,twoside,journal]{IEEEtran}
\IEEEoverridecommandlockouts
\usepackage{subfigure} 
\usepackage{graphicx}

\usepackage{amsmath,graphicx,amssymb,mathtools,bm}
\usepackage{subfigure}
\usepackage{hyperref}
\usepackage{cite}
\usepackage{amsmath,amssymb,amsfonts}
\usepackage{algorithmic}
\usepackage{graphicx}
\usepackage{textcomp}
\usepackage{xcolor}
\usepackage{verbatim}  
\usepackage{graphicx}  
\usepackage{bm}  
\usepackage{mathrsfs} 
\usepackage{algorithm} 
\usepackage{algorithmic} 
\usepackage{booktabs}
\usepackage{textcomp}  
\usepackage{multirow}  
\usepackage{lettrine}   
\usepackage{graphicx}  
\usepackage{color}  
\usepackage{amsmath}
\usepackage{amssymb}
\usepackage{stfloats}
\usepackage{caption}  
\usepackage{color}  

\UseRawInputEncoding
\usepackage[linesnumbered,ruled,vlined,algo2e]{algorithm2e}

\def\BibTeX{{\rm B\kern-.05em{\sc i\kern-.025em b}\kern-.08em
    T\kern-.1667em\lower.7ex\hbox{E}\kern-.125emX}}
\begin{document}
	\newcommand{\tabincell}[2]{\begin{tabular}{@{}#1@{}}#2\end{tabular}}
   \newtheorem{Property}{\it Property} 
  
 \newtheorem{Proposition}{\bf Proposition}
\newtheorem{remark}{Remark}
\newenvironment{Proof}{{\indent \it Proof:}}{\hfill $\blacksquare$\par}

\title{Flexible Antenna Arrays for \\ Wireless Communications: Modeling and Performance Evaluation
}

\author{Songjie Yang, Jiancheng An, \IEEEmembership{Member,~IEEE}, Yue Xiu, Wanting Lyu, Boyu Ning, \IEEEmembership{Member,~IEEE}, \\ Zhongpei Zhang, \IEEEmembership{Member,~IEEE},
	M\'erouane Debbah, \IEEEmembership{Fellow,~IEEE}, Chau Yuen, \IEEEmembership{Fellow,~IEEE}
	\vspace{-0.5cm}

\thanks{
	S. Yang, W. Lyu, B. Ning and Z. Zhang are with the National Key Laboratory of Wireless Communications, University of Electronic Science and Technology of China, Chengdu 611731, China. (e-mail:	yangsongjie@std.uestc.edu.cn; lyuwanting@yeat.net; boydning@outlook.com; zhangzp@uestc.edu.cn).
	
		X. Yue is with College of Air Traffic Management, Civil Aviation Flight University of China, Sichuan, China, 618311. (e-mail: xiuyue12345678@163.com).
		
	M. Debbah is with KU 6G Research Center, Khalifa University of Science and Technology, P O Box 127788, Abu Dhabi, UAE (email: merouane.debbah@ku.ac.ae).
	
J. An and C. Yuen are with the School of Electrical and Electronics Engineering, Nanyang Technological University, Singapore 639798 (e-mail: jiancheng\_an@163.com; chau.yuen@ntu.edu.sg). 
}}
\maketitle

\begin{abstract}
Flexible antenna arrays (FAAs), distinguished by their rotatable, bendable, and foldable properties, are extensively employed in flexible radio systems to achieve customized radiation patterns. This paper aims to illustrate that FAAs, capable of dynamically adjusting surface shapes, can enhance communication performances with both omni-directional and directional antenna patterns, in terms of multi-path channel power and channel angle Cram\'{e}r-Rao bounds. To this end, we develop a mathematical model that elucidates the impacts of the variations in antenna positions and orientations as the array transitions from a flat to a rotated, bent, and folded state, all contingent on the flexible degree-of-freedom.
Moreover, since the array shape adjustment operates across the entire beamspace, especially with directional patterns, we discuss the sum-rate in the multi-sector base station that covers the $360^\circ$ communication area. Particularly, to thoroughly explore the multi-sector sum-rate, we propose separate flexible precoding (SFP), joint flexible precoding (JFP), and semi-joint flexible precoding (SJFP), respectively.
In our numerical analysis comparing the optimized FAA to the fixed uniform planar array, we find that the bendable FAA achieves a remarkable $156\%$ sum-rate improvement compared to the fixed planar array in the case of JFP with the directional pattern. Furthermore, the rotatable FAA exhibits notably superior performance in SFP and SJFP cases with omni-directional patterns, with respective $35\%$ and $281\%$.
	
\end{abstract}
\section{Introduction}   

As wireless communication continues to advance, ongoing technological developments aim to explore new dimensions to achieve increased degrees-of-freedom (DoFs). A critical area of focus is the exploration of wireless channel characteristics, leveraging their properties to enhance communication and sensing capabilities. This encompasses strategies such as harnessing the sparsity of millimeter-wave/terahertz channels for beamspace signal processing \cite{mmw1,mmw2}, employing reconfigurable intelligent surfaces (RISs) to augment channel propagation \cite{RIS1,RIS2}, and exploring near-field spherical-wave channels to exploit the distance DoF \cite{NF1,NF2}. Despite these significant advancements, the potential of improving wireless channel conditions continues to be a vast and promising area for further research and exploration.

Recently, the emergence of movable antennas (MAs) marks a significant stride in wireless technologies. Specifically, by dynamically modifying the antenna's position, one could flexibly adjust the channel conditions to achieve constructive and destructive path interference and path orthogonalization  \cite{MA1,M3,MA2}. This dynamic intervention facilitates various outcomes, such as constructive and destructive path interference and path orthogonalization, which are integral to altering channel conditions. Parallel to this, fluid antenna systems (FASs) adopt a similar philosophy with antenna position optimization \cite{FAS1}. Both MA and FAS represent a synergistic alignment with earlier discussed techniques, reflecting a unified pursuit of impacting wireless channels for enhanced operational efficiency. In fact, it is not a new topic to investigate the impact of antenna position on communication rate and beam pattern optimization, such as antenna selection that selects certain subset of antenna positions given an antenna set to maximize communication performance \cite{AS1,AS2}, and array synthesis that optimizes antenna positions, excitation coefficients and even orientation to achieve the desired beam indicator such as maximizing the directivity and suppressing sidelobes \cite{SAS1}. All these four techniques regard antenna position as a flexible DoF, when applying to various applications. 

Beyond adjusting the position of individual antenna elements, recent research has delved into antenna shapes, particularly flexible antennas fabricated from diverse conductive materials and substrates, as thoroughly examined in \cite{FA1,FA2}, these antennas select substrates based on dielectric qualities and resistance to mechanical changes such as bending, twisting, and wrapping.
This flexibility adds a new dimension to deployment strategies, enabling adaptive positioning and orientation of elements. As a result, it facilitates the creation of versatile radiation patterns across various frequencies and angles, significantly enhancing adaptability and performance.
The flexible microstrip antenna, notable for its lightweight, low-profile, and simplistic structure, is extensively used in designing a range of conformal antennas. \cite{FA3} introduced a broadband flexible microstrip antenna array characterized by a combination of flexible substrate and honeycomb core layers. This configuration augmented both mechanical and electromagnetic properties, facilitating surface adaptability and load-bearing capacity. Demonstrating enhanced broadband functionality and diminished sidelobe levels under different bending and loading scenarios, this antenna is aptly suited for a variety of applications.

Extending the flexible element concept, the flexible substrate allows for the deformation of the array structure, giving rise to flexible antenna arrays (FAAs) \cite{FAA1,FAA2,FAA3,FAA4,FAA5}. Specifically, \cite{FAA2} explores the versatility of linear antenna arrays when wrapped around a cylindrical surface, assessing their irradiation performance in terms of side lobe level, mutual coupling, and other factors. \cite{FAA1} investigated a flexible array capable of bending in both horizontal and vertical planes, with concave and convex radii of less than 23 cm, while maintaining full functionality and programmability, including focusing, pattern recovery, and two-dimensional beam steering.
Except for bending structures, folding structures are considered the promising spaceborne devices among mechanical structures due to their distinctive transformable characteristics. Several works related to materials have investigated the integration of folding with communication devices, resulting in remarkable performance improvements, such as wideband switchable absorption \cite{fold2}. In particular, \cite{fold1} utilized folding techniques and flexible electronic materials in metamaterial design to achieve large-depth reflection modulation over an ultra-wideband by leveraging the folding DoF. It discussed the potential of this approach in extensive satellite communications due to its lightweight, foldability, and low-cost properties.
 Moreover, \cite{FAA5} introduced a metasurface adaptable to various dynamic shapes through the use of a Lorentz-force-driven mesh, facilitating easier shape manipulation. This study underscored the potential of integrating advanced materials and techniques in developing intelligent and shape-adaptive systems.
The above-mentioned works primarily focus on the antenna domain, examining FAAs from various perspectives such as radiation properties. In \cite{RIS2}, which pertains to wireless communications, the authors studied the rotatable RIS in the aerial environment to enhance MIMO communications. Their findings demonstrated significant performance enhancements facilitated by the rotating DoF.

On the other hand, multi-sector base stations (BS), sometimes referred to as multi-panel, are preferred in practical communication systems due to their ability to achieve full area coverage. \cite{MS2} discussed the possibility of deploying multiple antenna arrays at each BS and user equipment, enabling a sectorized deployment. It demonstrated that increasing the number of sectors can enhance the signal-to-interference-plus-noise ratio (SINR), thereby improving throughput and reducing latency, particularly for users with poor channel conditions.
In \cite{MS1}, a multi-sector BS was examined to evaluate cell communication performance. Notably, the authors investigated uniform circular arrays, shaped as a cyclinder with three panels, compared to three-panel uniform planar arrays (UPAs). Analysis of coverage tradeoffs, in terms of single-user achievable rates, revealed that with uniform user distributions, the cylindrical arrays provided higher minimum rates and less variance compared to sectorized UPAs. However, this was achieved at the cost of sacrificing some peak rates. Consequently, the cylindrical arrays may be considered suitable for applications prioritizing high reliability, such as industrial environments.
\cite{MS3} delved into sectorization for BS configuration, combining sparse arrays accounting for mutual coupling, pattern tapering, polarization, and multipath effects. Additionally, certain full-duplex works considered two-sector or two-panel BS \cite{FD1,FD2}, with one panel transmitting and the other receiving signals, aimed at suppressing self-interference caused by the dual-panel arrays.
Extending the concept of multi-sector BS, \cite{MS4} considered a multi-sector RIS, employing multiple sub-RISs to fully exploit the RIS's coverage enhancement capability in aiding wireless communications.

As of now, there is scarcely any research on FAAs with adaptive shapes in wireless communications. Given the diverse communication scenarios and varying channel conditions, FAAs hold promise for enhancing communication performance through added DoFs by dynamically optimizing array shape, such as providing enhanced coverage for multi-sector communications. 
In this paper, we explore the potential of FAAs for enhancing communication performance. While there are various FAA shapes, we focus on three shapes: rotatable, bendable, and foldable FAAs, to model and evaluate their impact on communication performance through the flexible DoF. Our novelty and contributions are summarized as follows\footnote{The source code for this paper is openly available at \url{https://github.com/YyangSJ/Flexible_antenna_arrays_wireless_communications}}.

\begin{itemize}
	\item We examine how the antenna position and orientation mapping change as the FAA rotates, bends, and folds. To maintain consistency, we consider a single flexible DoF --- rotating angle, bending angle, or folding angle --- for all three types of FAAs. We provide mathematical models to determine the position and orientation of antennas as a function of this flexible DoF. This approach enables us to perceive the wireless channel's influence by the flexible DoF, introducing a new factor to enhance channel conditions. To assess the multi-path channel influenced by FAAs, we focus on two critical aspects related to channel properties: multi-path channel power and the channel Cram\'{e}r-Rao bound (CRB). We examine whether there is performance enhancement in these aspects for both omni-directional and directional antenna patterns. 
	\item We introduce three multi-sector BS schemes for flexible precoding, aiming to optimize both precoding (antenna coefficients) and flexible DoF (array shape) simultaneously to maximize multi-sector sum-rate.
	1) Separate flexible precoding (SFP): each FAA independently conducts flexible precoding for the users within its own sector.
	2) Joint flexible precoding (JFP): all FAAs collaboratively perform flexible precoding for all users simultaneously.
	3) Semi-Joint flexible precoding (SJFP): each FAA conducts precoding separately for the users within its own sector. Subsequently, all FAAs jointly optimize their flexible DoFs.
	\item 
	Based on the three sector processing schemes, we introduce Bayesian optimization-based zero-forcing (BO-ZF) for flexible precoding. This method leverages the closed-form expression of ZF precoding to formulate the sum-rate maximization problem concerning the flexible DoFs, followed by global optimization with BO.
\end{itemize}

The rest of this paper is organized as follows: Section \ref{SMD} describes the system model of the sectorized BS, including the multi-user signal model, multi-path channel model, and omnidirectional and directional antenna radiation pattern. Section \ref{FM} studies three types of shapes for FAAs impacting wireless communications: rotatable, bendable, and foldable.   In Section \ref{MUSR}, the multi-sector sum-rate is assessed in FAAs, followed by the BO solutions in Section \ref{BO_FAA}. Section \ref{SIMR} presents several numerical results for demonstrating the proposed FAAs' effectiveness. Finally, Section \ref{Con} summarizes this paper and offers potential future directions.
 
{\emph {Notations}}:
 ${\left(  \cdot  \right)}^{ H}$ and $\left(\cdot\right)^{-1}$ denote   conjugate transpose and inverse, respectively. $\vert\cdot\vert$ represents the modulus. $\Vert\mathbf{a}\Vert_2$ and
$\Vert\mathbf{A}\Vert_F$ denotes the $\l_2$ norm of vector $\mathbf{a}$ and Frobenius norm of matrix $\mathbf{A}$, respectively. $\rm Tr(\cdot)$ denotes the trace. $\odot$ represents the Hadamard product. $\Re\{\cdot\}$ denotes the real part of a complex-value number.
   $[\mathbf{A}]_{i,:}$ and $[\mathbf{A}]_{:,j}$ denote the $i$-th row and the $j$-th column of matrix $\mathbf{A}$, respectively.  $\mathbb{E}\{\cdot\}$ denotes the expectation.  $\mathbf{I}_K$ denotes a $K\times K$ identity matrix. Finally, $\mathcal{CN}(\mathbf{a},\mathbf{A})$ is the complex Gaussian distribution with mean $\mathbf{a}$ and covariance matrix $\mathbf{A}$.

\section{System Model}\label{SMD}
\begin{figure*}
	\centering 
	\includegraphics[width=6.9in]{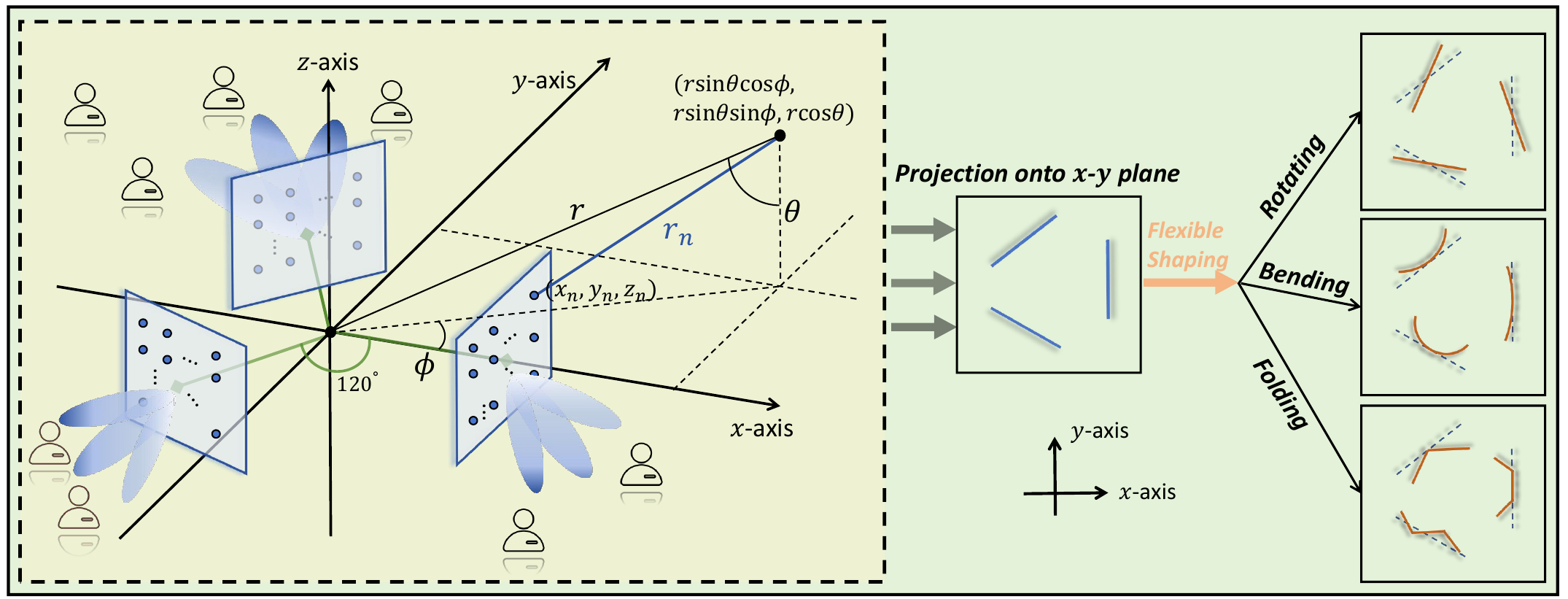}
	\caption{Multi-sector communication system and three FAA types.}\label{system} 
\end{figure*}

In our study, we concentrate on a multi-user downlink system where the sectorized $360^\circ$-coverage BS comprises $M\triangleq 3$ FAAs, each tasked with covering a sector of the surrounding communication sector. This arrangement is illustrated on the left in Fig. \ref{system}. One FAA is equipped with $N\triangleq N_h\times N_v$ antenna elements, where $N_h$ and $N_v$ denote the counts of horizontal and vertical antennas, respectively. These sectors have the dual capability of jointly processing all users' signals via a central processing unit, and also of independently handling the signals of users within their specific directional sectors. We assume there are $K$ users in each sector, all equipped with a single antenna and capable of receiving one data stream from the BS, making the total user count in the covered area $3K$.

Our system configuration facilitates adaptive adjustments of each FAA's shape, leveraging flexible substrates or other advanced flexible antenna technologies mentioned earlier. Specifically, we investigate three types of FAA designed for straightforward implementation: those with rotatable, bendable, and foldable shapes, as depicted on the right in Fig. \ref{system}. It should be noted that the flexibility is only considered to be reflected in horizontal directions ($x$-$y$ plane).

\subsection{Signal Model}
In the downlink communication system, the received signal $	u_{m^\prime_k}$ at the $k$-th user of the $m^\prime$-th sector, $k\in\{1,\cdots,K\}$, $m^\prime\in\{1,2,3\}$, can be expressed by
\begin{equation}
	u_{m^\prime_k}=\sum_{m=1}^{3}\mathbf{h}_{m,m^\prime_k}^H\mathbf{F}_m\mathbf{s}_m+n_{m_k},
\end{equation}
where $\mathbf{h}_{m,m^\prime_k}\in\mathbb{C}^{N\times 1}$ is the channel between the $m$-th FAA and the $k$-th user of the $m^\prime$-th sector, $\mathbf{F}_m\triangleq[\mathbf{f}_{m_1},\cdots,\mathbf{f}_{m_K}]\in\mathbb{C}^{N\times K}$ is the precoding matrix of the $m$-th FAA,
$\mathbf{s}_m\in\mathbb{C}^{K\times 1}$ represents the $K$ data streams for $K$ users in the $m$-th sector, $n_{m_k}$ represents the additive white Gaussian noise following $\mathcal{CN}(0,\sigma^2_n)$.

It is important to note that the interference level at user $m_k$ relies on the system's processing technique.

In the context of sector-separate processing, all signals other than $\mathbf{f}_{m_k}$ are considered interference to user $m_k$. Assuming i.i.d. transmit data, the sector-separate SINR for the user $m_k$ can be expressed as follows:

\begin{equation}\label{SINRS0}
	\begin{aligned}
		&	{\rm SINR}_{m_k}= \\ &\frac{\left\vert{\mathbf{h}}_{m,m_k}^H\mathbf{f}_{m_k}\right\vert^2}{\sum_{i,i\neq k}^{K}\left\vert{\mathbf{h}}_{m,m_k}^H\mathbf{f}_{m_i}\right\vert^2+\sum_{m^\prime\neq m}^3 \sum_{i}^{K}\left\vert{\mathbf{h}}_{m^\prime,m_k}^H\mathbf{f}_{m^\prime_i}\right\vert^2+\sigma_n^2}.
	\end{aligned} 
\end{equation}

In sector-joint processing, all FAAs collaborate to perform precoding for each user. Similarly, the sector-joint SINR for the $m_k$-th user can be defined as follows:
\begin{equation} \label{SINRJ0}
	\begin{aligned}
		{\rm SINR}_{m_k}=\frac{\sum_{m=1}^3\left\vert{\mathbf{h}}_{m,m_k}^H\mathbf{f}_{m_k}\right\vert^2}{\sum_{m^\prime\neq m}^3 \sum_{i\neq k}^{K}\left\vert{\mathbf{h}}_{m^\prime,m_k}^H\mathbf{f}_{m^\prime_i}\right\vert^2+\sigma_n^2}.
	\end{aligned}
\end{equation}

\subsection{Channel Model}\label{CM}
We foucus on the multipath channel, by assuming $L$ paths for all user channels, the channel of user $m_k$ is established by, for $\forall k,m,m^\prime$:
\begin{equation}\label{hk}
	\begin{aligned}
	&	\mathbf{h}_{m^\prime,m_k}=\\ &\sqrt{\frac{1}{L}}\sum_{l=1}^{L}\beta_{m^\prime,m_k,l} A_E(\theta_{m^\prime,m_k,l},\phi_{m^\prime,m_k,l})\mathbf{g}(\theta_{m^\prime,m_k,l},\phi_{m^\prime,m_k,l}),
	\end{aligned}
\end{equation}
where $\beta_{m^\prime,m_k,l}$ is the complex path gain of the $l$-th path of the channel from the $m^\prime$-th FAA to the $m_k$-user, $L$ is number of spatial channel paths,
$\phi_{m^\prime,{m_k},l}$ and $\theta_{m^\prime,{m_k},l}$ denote the azimuth/elevation angles of the $l$-th path of the ${m_k}$-th user channel regarding to the $m^\prime$-th FAA, $A_E(\theta,\phi)$ represents the antenna element radiation coefficient, assumed here that the patterns of all antennas are identical, 
and the far-field array manifold $\mathbf{g}(\theta_{m^\prime,{m_k},l},\phi_{m^\prime,{m_k},l})\in\mathbb{C}^{N\times 1}$ follows:
\begin{equation}\label{gn}
[\mathbf{g}(\theta,\phi)]_n\approx e^{-j\frac{2\pi}{\lambda}( x_n\sin\theta\cos\phi+y_n\sin\theta\sin\phi+z_n\cos\theta) }.
\end{equation}

   The channel $\mathbf{H}_{m,m^\prime}\in\mathbb{C}^{N\times K}$ between the $m$-th FAA and the users in the $m^\prime$-th sector  is defined by
\begin{equation}
	\mathbf{H}_{m,m^\prime}\triangleq\left[\overline{\mathbf{h}}_{m, {m^\prime}_1} , \overline{\mathbf{h}}_{m,{m^\prime}_2} ,\cdots,\overline{\mathbf{h}}_{m,{m^\prime}_K} \right].
\end{equation} 
Then, the whole channel matrix $\overline{\mathbf{H}}\in\mathbb{C}^{3N\times 3K}$ is given by
\begin{equation}\label{dH}
	\overline{\mathbf{H}}=
	\begin{bmatrix}
		\mathbf{H}_{1,1}&	\mathbf{H}_{1,2}&	\mathbf{H}_{1,3} \\
		\mathbf{H}_{2,1}&	\mathbf{H}_{2,2}&	\mathbf{H}_{2,3}\\
		\mathbf{H}_{3,1}&	\mathbf{H}_{3,2}&	\mathbf{H}_{3,3}
	\end{bmatrix}.
\end{equation} 
\subsection{Antenna Radiation Pattern}\label{ARP}
The spatial channel model typically used in Eqn. (\ref{hk}) often assumes an omnidirectional antenna with a constant $A_E(\theta,\phi)\equiv 1$ across the entire beamspace. However, the use of directional antennas is gaining prominence in wireless communications, as suggested in \cite{3GPP2}. Indeed, directional antennas are commonly preferred in practical engineering systems due to their advantages in focusing energy and improving signal quality.

In this study, both omnidirectional and directional antenna patterns are considered, with the directional antenna specifically accounting for the influence of antenna orientation. Regarding the omnidirectional pattern, we have:
\begin{equation}
	G_E(\theta,\phi)=1, \ \theta\in [0,\pi], \phi \in [0,2\pi].
\end{equation}
The relationship between the pattern coefficient $A_E(\theta,\phi)$ and the pattern radiation power $G_E(\theta,\phi)$ is given by $A_E(\theta,\phi)=\sqrt{G_E(\theta,\phi)}$.

For the directional pattern, this paper considers the cosine pattern, widely utilized in the antenna community \cite{cos1,cos2}. However, since the commonly used form assumes a $z$-axis orientation, we modify the pattern function to accommodate our $x$-axis orientation case while retaining the similar properties. This modified pattern function is presented as:
\begin{equation}\label{GE}
	G_E(\theta,\phi)=
		\begin{cases}
G \sin^\kappa \theta \cos^\kappa \phi,& \theta\in[0,\pi],\phi\in [-\frac{\pi}{2},\frac{\pi}{2}], \\
	0, & \text{otherwise},
	\end{cases}
\end{equation}
where $\kappa\geq 1$ is the pattern sharpness factor. A larger value of $\kappa$ corresponds to a radiation pattern having stronger directivity. The power normalization factor $G$ is defined such that $\int_{\Omega} G \sin^\kappa \theta \cos^\kappa \phi {\rm d} \Omega = 4\pi $, where $\Omega$ represents the spherical space. In this case, $G=2(1+\kappa)$ is derived as follows.

The mathematical expression of the widely used cosine pattern orienting $z$-axis is given by \cite{cos1,cos2}
\begin{equation}
G_E(\theta,\phi)=\begin{cases}
	G\cos^\kappa \theta, & \theta\in [0,\frac{\pi}{2}], \phi \in [0,2\pi] \\
	0, & \text{otherwise}
\end{cases},
\end{equation}
where $G\triangleq 2(\kappa+1)$ for normalization has been proved in \cite{cos1}.
To apply it in our $x$-axis orientation case, the pattern response is changed to $\sin\theta\cos\theta$, and incurs in Eqn. (\ref{GE}).

\section{Three FAA Models}\label{FM}
\begin{figure*}
	\centering 
	\includegraphics[width=6.32in]{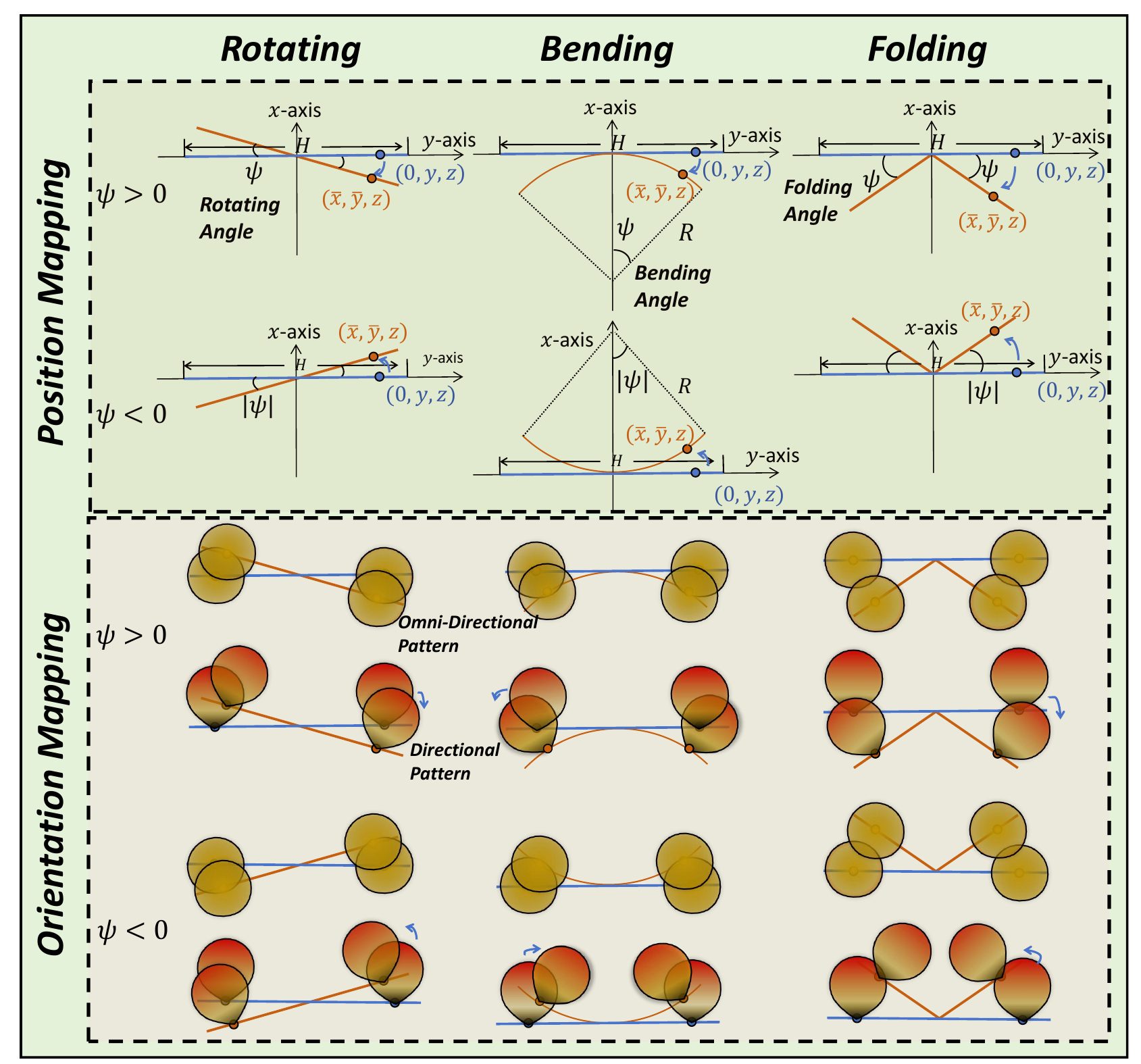}
	\caption{Antenna position mapping and orientation mapping of three FAA models.}\label{mapping} 
\end{figure*}

This section investigates the impact of flexible DoFs on antenna positions, orientations, and wireless channels. To maintain simplicity and clarity, all three shapes employ only a single DoF: the rotating angle denoted as $\psi$ for rotatable FAAs, and the bending angle and the folding angle, also denoted as $\psi$, for bendable and foldable FAAs, respectively. This concept is illustrated in Fig. \ref{mapping}.
  \subsection{Flexible Rotating Model}\label{FRM}
We first consider a rotatable array that is capable of rotating in the horizontal dimension. We aim to model the variations in antenna positions and orientations.
\subsubsection{Antenna Position Mapping}
As the array rotates around the $z$-axis, the $z$-coordinate remains unchanged. This rotation process is depicted on the top-left of Fig. \ref{mapping}. Moreover, this rotation can be mathematically represented by the rotation matrix $\mathbf{R}_{xy}$ given by
\begin{equation}\label{rot_m}
\mathbf{R}_{xy}(\psi)=\begin{bmatrix}
 \cos\psi &-\sin\psi \\
 \sin\psi &\cos\psi
\end{bmatrix},
\end{equation}
where $\psi$ is the rotation angle, measured in radians, with the positive value denoting counterclockwise rotation and the negative value denoting clockwise rotation. 

Now, it is easy to get the $\psi$-rotated antenna position by
\begin{equation}
	\begin{bmatrix}
		\overline{x} \\ \overline{y}
	\end{bmatrix} = \mathbf{R}_{xy}(\psi)	\begin{bmatrix}
	{x} \\ {y}
\end{bmatrix}.
\end{equation}

Assuming the planar array is deployed on the $y$-$z$ plane with half-wavelength inter-element spacing, the rotated coordinates are as follows:
\begin{equation}
	\begin{cases}
		\overline{x}_{n_h}=- \frac{2n_h-N_h-1}{2}d \sin\psi, \ n_h\in\{1,\cdots,N_h\}, \\
		\overline{y}_{n_h}= \frac{2n_h-N_h-1}{2}d \cos\psi, \ n_h\in\{1,\cdots,N_h\}, \\
		\overline{z}_{n_v}=  \frac{2n_v-N_v-1}{2}d , \ n_v\in\{1,\cdots,N_v \}.
	\end{cases}
\end{equation} 
\vspace{-0.1cm}
\subsubsection{Antenna Orientation Mapping}

The rotatable array structure not only alters the position of the antennas but also their orientation, thereby significantly influencing the radiation pattern of the antenna elements.
We notice that as the array rotates, all elements experience uniform rotation, leading to consistent alterations in the power distribution of all element patterns by the rotating angle $\psi$, represented as $A_E(\theta,\phi-\psi)$, as depicted in the bottom-left of Fig. \ref{mapping}.
By contrast, when examining the omnidirectional pattern, the radiation energy remains unchanged by the antenna orientation.
 
\subsection{Flexible Bending Model}\label{FBM}
Secondly, we consider the bendable array that is capable of bending in the horizontal dimension by characterizing the variations in antenna positions and orientations.
\subsubsection{Antenna Position Mapping}

The planar array structure using the parametrization can be represented by
\begin{equation}
	\begin{cases}
		x=0,  \\
		y=\frac{H}{2}t,\ t\in [-1,1],\\
		z=z_0,\ z_0\in [-V/2,V/2],
	\end{cases}
\end{equation}
where $V\triangleq (N_v-1)d$ with $d\triangleq \lambda/2$ and $H\triangleq (N_h-1)d$ are the vertical and horizontal array size, respectively. 

Denoting the bending angle as $\psi$, we can achieve the parametrization of an arc through the projection in the vertical direction, depicted at the top-center in Fig. \ref{mapping}, described by $	\overline{x}=R\left(\cos\left(\overline{t}\right)-1\right),
\overline{y}=R\sin\left(\overline{t}\right)$, $\overline{t}\in [-\psi,\psi]$, where $R$ denotes the radius of the bent arc. Note that when $\psi>0$, the arc bends toward the negative $x$-axis, as illustrated in the first case at the top-center in Fig. \ref{mapping}. Conversely, for $\psi<0$, the arc bends toward the $x$-axis, as depicted in the second case. Both those two cases share the same mathematical model.
 Combining the vertical coordinate, we can obtain the parametrization of the bent array:
\begin{equation} \label{XYZ}
	\begin{cases}
		\overline{x}=R\left(\cos\left(\overline{t}\right)-1\right),\ \overline{t}\in [-\psi,\psi], \\
		\overline{y}=R\sin\left(\overline{t}\right),\ \overline{t}\in [-\psi,\psi],\\
		\overline{z}=z_0,\ z_0\in [-V/2,V/2].
	\end{cases}
\end{equation}

Noticing $\frac{H}{2}=R\psi$, $\overline{t}=\psi t$, and $y=Ht$,
we derive the relationship between the original and the bent coordinates:
\begin{equation}\label{bxyz2}
	\begin{cases}
		\overline{x}=\frac{H}{2\psi}\left(\cos\left(\frac{2\psi}{H}y\right)-1\right), \ y\in[-H/2,H/2], \\
		\overline{y}=\frac{H}{2\psi}\sin\left(\frac{2\psi}{H}y\right), \ y\in[-H/2,H/2], \\
		\overline{z}=z,\ z\in [-V/2,V/2].
	\end{cases}
\end{equation}

Notably, with the discrete sampling, such as half-wavelength inter-element spacing $d$ for $y_{n_h}$, $n_h\in\{1,\cdots,N_h\}$, $\frac{2\psi}{H}y_n$ can be simplified into 
\begin{equation}\label{pnh}
	\begin{aligned}
		\frac{2\psi}{H}y_{n_h}=&\frac{2\psi}{(N_h-1)d}\left(-\frac{N_h-1}{2}+(n_h-1)\right)d\\
		=& -\psi +2\psi\frac{(n_h-1)}{N_h-1} \\
	\triangleq	& \psi_{n_h}.
	\end{aligned}
\end{equation}

Hence, the $N_h\times N_v$ FAA's antenna coordinates with respect to $\psi$ can be expressed by
\begin{equation}\label{bxyz3}
	\begin{cases}
		\overline{x}_{n_h}=R\left(\cos\left(\psi_{n_h}\right)-1\right), \ n_h\in\{1,\cdots,N_h\}, \\
		\overline{y}_{n_h}=R\sin\left(\psi_{n_h}\right), \ n_h\in\{1,\cdots,N_h\}, \\
		\overline{z}_{n_v}=  \frac{2n_v-N_v-1}{2}d , \ n_v\in\{1,\cdots,N_v \},
	\end{cases}
\end{equation} 
where $R=\frac{(N_h-1)d}{2\psi}$.
\subsubsection{Antenna Orientation Mapping}

The bendable array structure not only alters the position of the antennas but also their orientation, which significantly influences the radiation pattern of the antenna elements. For the directional antenna case, the elements are initially assumed to be positioned on the $y$-$z$ plane and oriented along the $x$-axis. However, with the bending of the array shape, the antennas may orient in different directions. To accurately represent this, we postulate that the antenna orientation (radiation pattern) correlates with the normal to the array surface post-bending, as illustrated at the bottom-center in Fig. \ref{mapping}.
 
Given that bending is solely considered in the horizontal dimension, we can derive the element radiation pattern, where the azimuth angle is influenced by the bending angle $\psi$: $
	A_E(\theta,\phi-\psi_{n_h})
$,
where $\psi_{n_h}$ is defined by Eqn. (\ref{pnh}). 
 
Following that, we can determine the radiation pattern for all elements and encapsulate them in a pattern matrix
$\mathbf{E}(\theta,\phi,\psi)\in\mathbb{C}^{N_h\times N_v}$, which is expressed as
\begin{equation}\label{E}
	\mathbf{E}(\theta,\phi,\psi)\triangleq\begin{bmatrix}
		A_E(\theta,\phi-\psi_{1})& \cdots&	A_E(\theta,\phi-\psi_{1})\\	A_E(\theta,\phi-\psi_{2})& 	 \cdots&	A_E(\theta,\phi-\psi_{2})\\  \vdots&\ddots&\vdots\\ 	A_E(\theta,\phi-\psi_{N_h}) &\cdots&A_E(\theta,\phi-\psi_{N_h})
	\end{bmatrix}.
\end{equation} 
 
 \vspace{-0.6cm}
\subsection{Flexible Folding Model}\label{FFM}

We next consider the folding FAA, which is capable of generating myriad shape alterations. In order to simplify the analysis, we assume only one fold line centered around the array in the horizontal dimension.

\subsubsection{Antenna Position Mapping}
In this context, the flexible folding model is similar to the rotating model, but with different rotating direction for the bi-fold sides. We divide the foldable array into two sides, the folding coordinates, based on the rotation matrix in Eqn. (\ref{rot_m}), are given by
\begin{equation}
	\begin{bmatrix}
		\overline{x} \\ \overline{y}
	\end{bmatrix} = \begin{cases}
\mathbf{R}_{xy}(\psi)	\begin{bmatrix}
	{x} \\ {y}
\end{bmatrix}, & \{x,y\}\in \text{side 1},\\
\mathbf{R}_{xy}(-\psi)	\begin{bmatrix}
	{x} \\ {y}
\end{bmatrix},  & \{x,y\}\in \text{side 2}.
\end{cases}
\end{equation}

Assuming the planar array is deployed on the $y$-$z$ plane with half-wavelength inter-element spacing, the folded coordinates are as follows:
\begin{equation}
	\begin{cases}
		\overline{x}_{n_h}=- \left\vert \frac{2n_h-N_h-1}{2} \right\vert d   \sin\psi , \ n_h\in\{1,\cdots,N_h\}, \\
		\overline{y}_{n_h}= \frac{2n_h-N_h-1}{2}d \cos\psi, \ n_h\in\{1,\cdots,N_h\}, \\
		\overline{z}_{n_v}=  \frac{2n_v-N_v-1}{2}d , \ n_v\in\{1,\cdots,N_v \}.
	\end{cases}
\end{equation} 

As depicted in the top-right of Fig. \ref{mapping}. Here, positive $\psi$ induces folding along the negative $x$-axis, while negative $\psi$ leads to folding along the $x$-axis. This is also evident in the mathematical expression of $\overline{x}$-mapping: $\overline{x}_{n_h}=- \left\vert \frac{2n_h-N_h-1}{2} \right\vert d \sin\psi$, which is an odd function with respect to $\psi$.

\subsubsection{Antenna Orientation Mapping}

As the array bi-folds, the elements on the left side or the right side undergo identical rotations, leading to two alterations in the power distribution of all element patterns by the folding angle $\psi$. The elements on the left side have $A_E(\theta,\phi+\psi)$, while those on the right side have $A_E(\theta,\phi-\psi)$.
\vspace{-0.4cm}

\subsection{Flexible Channel Mapping}
Upon revisiting the channel model in Eqn. (\ref{hk}) and considering the variations of antenna positions and orientations, the channel $\overline{\mathbf{h}}_k$, which relies on the flexible DoF $\psi$, representing the rotating angle in Section \ref{FRM}, the bending angle in Section \ref{FBM}, or the folding angle in Section \ref{FFM}, can be expressed as:
\begin{equation}\label{hk2}
	\begin{aligned}
		 	\overline{\mathbf{h}}_{m^\prime,m_k}(\psi)= \sqrt{\frac{1}{L}}  \sum_{l=1}^{L}&\beta_{m^\prime,m_k,l}  \mathbf{e}(\theta_{m^\prime,m_k,l},\phi_{m^\prime,m_k,l},\psi) \\ 
		 	&\odot\mathbf{g}(\theta_{m^\prime,m_k,l},\phi_{m^\prime,m_k,l},\psi),
	\end{aligned}
\end{equation}
where $\mathbf{e}(\theta,\phi,\psi)\triangleq {\rm vec}\left(\mathbf{E}(\theta,\phi,\psi)\right)\in\mathbb{C}^{N_hN_v\times 1}$ comprises the radiation patterns of all elements, representing the column-stacking of Eqn. (\ref{E}). Meanwhile, $\mathbf{g}(\theta,\phi,\psi)$ denotes the planar-wave array manifold from Eqn. (\ref{gn}), with flexible position mapping by $\psi$.
Whether each of the three FAA is adopted, the value of $\psi$ can be adjusted to customize a favorable channel for enhancing wireless channel performance in data transmission.  
 
Building upon the above analysis, we now assess channel power and channel CRB, which are crucial metrics in wireless communications, to shed light on the impact of the FAAs.

\subsubsection{Multi-Path Channel Power}

The channel power plays a crucial role in determining the received signal-to-noise ratio (SNR), as it is influenced by both the path gain and angle information within the far-field region. Specifically, in the scenario where one FAA serves a single-antenna user, we can simplify the channel related to $\psi$ as follows:
\begin{equation}
		\overline{\mathbf{h}}(\psi)=\sqrt{\frac{1}{L}}\sum_{l=1}^{L}\beta_{l} \mathbf{e}(\theta_{l},\phi_{l},\psi) \odot \mathbf{g}(\theta_{l},\phi_{l},\psi). 
\end{equation}

Following that, the channel power is calculated by
\begin{equation}\label{hp}
	\begin{aligned}
		&\left\vert\overline{\mathbf{h}}^H(\psi)\overline{\mathbf{h}}(\psi)\right\vert^2\\
		= &\frac{1}{L}\sum_{l=1}^{L} \vert \beta_l\vert^2\left\Vert \mathbf{e}(\theta_l,\phi_l,\psi) \odot \mathbf{g}(\theta_l,\phi_l,\psi)  \right\Vert_2^2+ \\ &
		\frac{1}{L} \sum_{l_1=1}^{L} \sum_{l_2\neq l_1}^L 2\Re\{\beta_{l_2}^*\beta_{l_1}
\left(  \mathbf{e}(\theta_{l_2},\phi_{l_2},\psi) \odot \mathbf{g}(\theta_{l_2},\phi_{l_2},\psi)  \right)^H \\ & \times \left(  \mathbf{e}(\theta_{l_1},\phi_{l_1},\psi) \odot \mathbf{g}(\theta_{l_1},\phi_{l_1},\psi)  \right)\}
+\\
		=& \frac{1}{L}\sum_{l=1}^{L} \vert \beta_l\vert^2\left\Vert \mathbf{e}(\theta_l,\phi_l,\psi) \odot \mathbf{g}(\theta_l,\phi_l,\psi)  \right\Vert_2^2+ \\ & \frac{1}{L} \sum_{l_1=1}^{L} \sum_{l_2> l_1}^L2\Re\{  \beta_{l_1}\beta^*_{l_2}
		\left(\mathbf{e}(\theta_{l_2},\phi_{l_2},\psi) \odot\mathbf{e}(\theta_{l_1},\phi_{l_1},\psi)\right)^T \\
		&\times \left(\mathbf{g}^*(\theta_{l_2},\phi_{l_2},\psi) \odot\mathbf{g}(\theta_{l_1},\phi_{l_1},\psi)\right) \} .
	\end{aligned}
\end{equation} 

 \begin{figure}
	\centering 
	\includegraphics[width=3.6in]{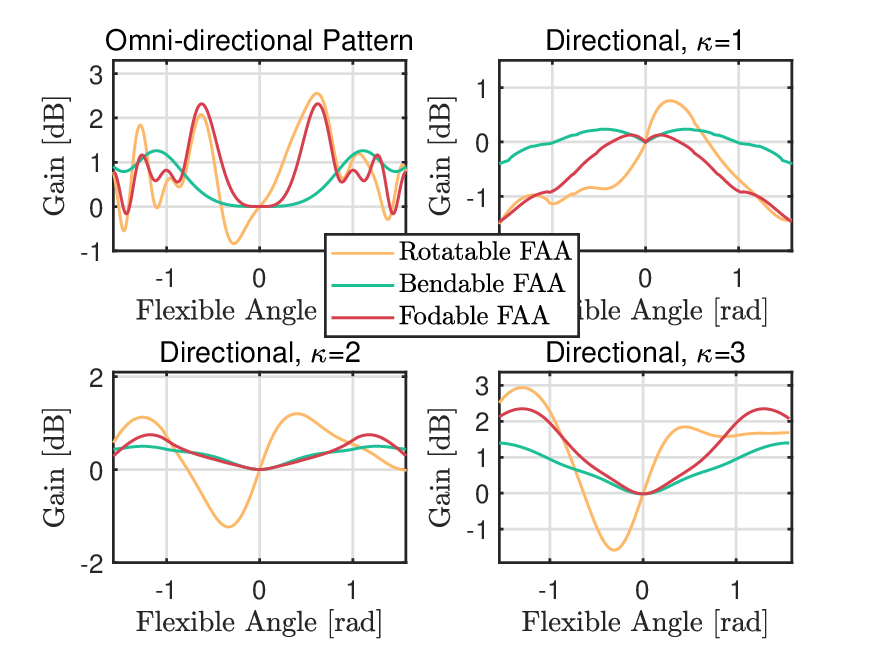}
	\caption{Channel power gain of the three FAAs over the fixed array.}\label{power_plot} 
\end{figure}

The equation above characterizes how the flexible DoF $\psi$ influences the channel power.
To illustrate the impacts of different FAA models, we consider an $8 \times 2$ FAA setup with $\lambda=0.03$ meters for a single user experiencing $5$ channel paths, where $\bm{\theta}=[\frac{\pi}{2}, \frac{2\pi}{3}, \frac{\pi}{6}, \frac{\pi}{3}, \frac{\pi}{2}]$, $\bm{\phi}=[-\frac{\pi}{2},-\frac{\pi}{3},\frac{\pi}{4},\frac{\pi}{6},\frac{\pi}{2}]$, and $\beta_l=1$ for all $l$. The flexible DoF values range from $[-\frac{\pi}{2},\frac{\pi}{2}]$. In this scenario, the ratio of channel power of the flexible channel to the fixed channel concerning the flexible DoF (flexible angle $\psi$) is depicted in Fig. \ref{power_plot} under both the omni-directional and directional patterns with $\kappa=1$.
We observe that under the omni-directional pattern, the rotatable FAA achieves approximately a $2.5$ dB gain at around $\psi=0.7$, where $0$ dB represents the fixed UPA, corresponding to $\psi=0$. Conversely, under the directional pattern, as $\kappa$ increases, indicating higher directivity strengths, this gain becomes more pronounced.

\subsubsection{Multi-Path Channel CRB}

The other significant case involves spatial channel estimation error, which depends on the channel condition. Some studies have applied the angle CRB to measure this error. Notably, the spatial channel with directional antenna patterns has not been addressed yet. Therefore, focusing on our flexible FAA case with adjustable antenna positions and orientations, we derive the directional channel CRB to explore how the bending angle impacts it. In practice, it is meaningful to evaluate this to achieve adaptive beamforming for pursuing high estimation performance. For example, we could estimate coarse channel parameters, including path angles and gains, and then utilize them to optimize certain error metrics, such as CRB minimization (this metric is typically used for radar sensing and integrated sensing and communication (ISAC) \cite{CM,ISAC1}), with respect to the transmit pilot and the flexible DoF, in order to perform refined channel estimation procedures. Hence, in the following sections, we explore the relationship between the directional channel CRB and the flexible DoF.
 
Consider the uplink training model $\mathbf{u}=\overline{\mathbf{h}}(\psi)+\mathbf{n}$, where $\mathbf{u}\in\mathbb{C}^{N\times 1}$ and $\mathbf{n}\in\mathbb{C}^{N\times 1}$ denote the received pilot signal and the i.i.d. Gaussian noise vector with $\mathcal{CN}(0,\sigma_n^2\mathbf{I}_N)$, repsecitvely.

First, the channel parameter set $\bm{\xi}\in\mathbb{C}^{4L\times 1}$ inherent in $\overline{\mathbf{h}}$ is defined by
\begin{equation}
	\bm{\xi}\triangleq [\bm{\theta}^T,\bm{\phi}^T,\bm{\beta}_R^T,\bm{\beta}_I^T]^T,
\end{equation}
where $\bm{\theta}\triangleq [\theta_1,\cdots,\theta_L]\in\mathbb{C}^{L\times 1}$, $\bm{\phi}\triangleq [\phi_1,\cdots,\phi_L]\in\mathbb{C}^{L\times 1}$, $\bm{\theta}\triangleq [\beta_{R,1},\cdots,\beta_{R,L}]\in\mathbb{C}^{L\times 1}$, and $\bm{\theta}\triangleq [\beta_{I,1},\cdots,\beta_{I,L}]\in\mathbb{C}^{L\times 1}$ represent the path elevation angle, azimuth angle, gain's real part, and gain's imaginary part, respectively.

The CRB of the $l$-th parameter, $l\in\{1,\cdots,4L\}$, in $\bm{\xi}$ is given by
\begin{equation}\label{CRBX}
	\text{CRB}(\xi_l)= \left[ \bm{\mathcal{F}}^{-1}\right]_{l,l},
\end{equation}
where the fisher matrix $\bm{\mathcal{F}}$ is written as
\begin{equation}
	\begin{aligned}
		\bm{\mathcal{F}}\triangleq 
		\begin{bmatrix}
			\mathbf{J}_{\bm{\theta},\bm{\theta}}& \mathbf{J}_{\bm{\theta},\bm{\phi}} &\mathbf{J}_{\bm{\theta},\bm{\beta}_{R}} &\mathbf{J}_{\bm{\theta},\bm{\beta}_{I}} \\ 	\mathbf{J}^T_{\bm{\theta},\bm{\phi}}& \mathbf{J}_{\bm{\phi},\bm{\phi}} &\mathbf{J}_{\bm{\phi},\bm{\beta}_{R}} &\mathbf{J}_{\bm{\phi},\bm{\beta}_{I}} \\ 	\mathbf{J}^T_{\bm{\theta},\bm{\beta}_R}& \mathbf{J}^T_{\bm{\phi},\bm{\beta}_R} &\mathbf{J}_{\bm{\beta}_R,\bm{\beta}_{R}} &\mathbf{J}_{\bm{\beta}_R,\bm{\beta}_{I}} \\ 	\mathbf{J}^T_{\bm{\theta},\bm{\beta}_I}& \mathbf{J}^T_{\bm{\phi},\bm{\beta}_I} &\mathbf{J}^T_{\bm{\beta}_R,\bm{\beta}_{I}} &\mathbf{J}_{\bm{\beta}_I,\bm{\beta}_{I}} 
		\end{bmatrix} \in\mathbb{R}^{4L\times 4L}
	\end{aligned}
\end{equation}

According to the i.i.d. Gaussian noise-based CRB, we have
\begin{equation}
\mathbf{J}_{\bm{\theta},\bm{\phi}}=\frac{2}{\sigma_n^2} \Re\left\{ \left(\frac{\partial \overline{\mathbf{h}}(\psi) }{\partial\bm{\theta}} \right)^H\frac{\partial \overline{\mathbf{h}}(\psi) }{\partial\bm{\phi}}
\right\}\in\mathbb{C}^{L\times L},
\end{equation}
and $\mathbf{J}_{\bm{\theta},\bm{\theta}},\mathbf{J}_{\bm{\theta},\bm{\beta}_R}$, $ \mathbf{J}_{\bm{\theta},\bm{\beta}_I},\mathbf{J}_{\bm{\phi},\bm{\beta}_R}$, $ \mathbf{J}_{\bm{\phi},\bm{\phi}}$, $ \mathbf{J}_{\bm{\phi},\bm{\beta}_I}$, $\mathbf{J}_{\bm{\beta}_R,\bm{\beta}_I}$, $\mathbf{J}_{\bm{\beta}_R,\bm{\beta}_R}$, $\mathbf{J}_{\bm{\beta}_I,\bm{\beta}_I}$ have similar forms. The derivative functions are provided for all $\forall l\in\{1,\cdots,L\}$:
\begin{equation}
	\begin{aligned}
		\frac{\partial \overline{\mathbf{h}}(\psi)}{\partial {\theta}_l}=\sqrt{\frac{1}{L}} \beta_l &\left(\frac{\partial \mathbf{e}(\theta_l,\phi_l,\psi)}{\partial \theta_l}\odot \mathbf{g}(\theta_l,\phi_l,\psi)  \right. \\ &\left. + \frac{\partial \mathbf{g}(\theta_l,\phi_l,\psi)}{\partial \theta_l}\odot \mathbf{e}(\theta_l,\phi_l,\psi)
		\right) ,
	\end{aligned}
\end{equation} 
\begin{equation}
	\begin{aligned}
		\frac{\partial \overline{\mathbf{h}}(\psi)}{\partial {\phi}_l}=\sqrt{\frac{1}{L}} \beta_l &\left(\frac{\partial \mathbf{e}(\theta_l,\phi_l,\psi)}{\partial \phi_l}\odot \mathbf{g}(\theta_l,\phi_l,\psi)  \right. \\ &\left. + \frac{\partial \mathbf{g}(\theta_l,\phi_l,\psi)}{\partial \phi_l}\odot \mathbf{e}(\theta_l,\phi_l,\psi)
		\right) ,
	\end{aligned}
\end{equation} 
\begin{equation} 
	\frac{\partial \overline{\mathbf{h}} (\psi)}{\partial \beta_{R,l}}=\sqrt{\frac{1}{L}}\mathbf{e}(\theta_{l},\phi_{l},\psi) \odot \mathbf{g}(\theta_{l},\phi_{l},\psi),
\end{equation} 
\begin{equation} 
	\frac{\partial \overline{\mathbf{h}} (\psi)}{\partial \beta_{I,l}}=j\sqrt{\frac{1}{L}}\mathbf{e}(\theta_{l},\phi_{l},\psi) \odot \mathbf{g}(\theta_{l},\phi_{l},\psi).
\end{equation}

Finally, the CRB with respect to $\psi$ can be derived by Eqn. (\ref{CRBX}).

\begin{figure}
 	\centering 
 	\includegraphics[width=2.8in]{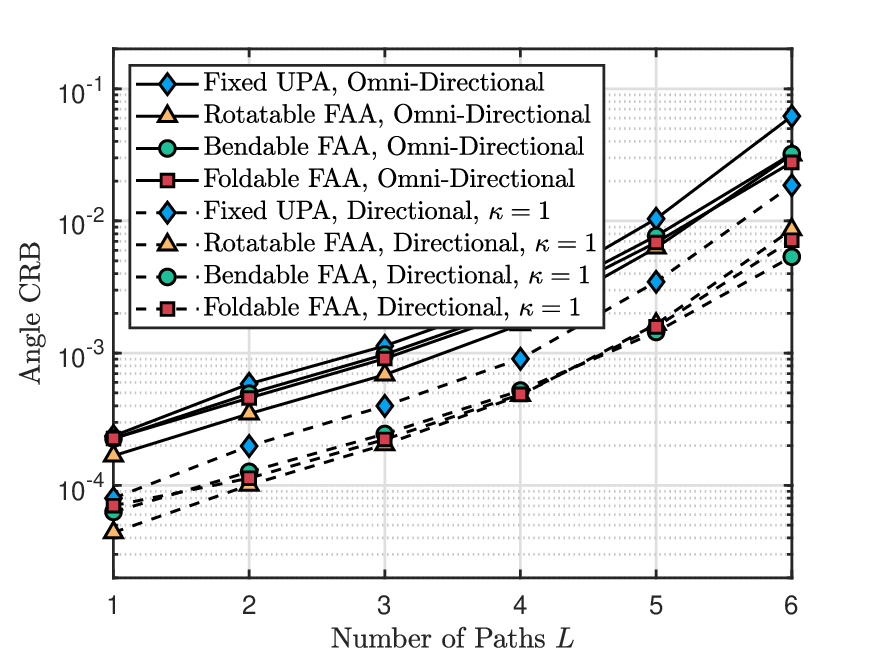}
 	\caption{Angle CRB versus the number of paths $L$ for three FAA and fixed UPA models under omni-directional and directional cosine pattern with $\kappa=1$.}\label{crb_plot} 
 \end{figure}

The $\psi$-related directional channel CRB influences channel estimation accuracy. Illustrated in Fig. \ref{crb_plot}, we present the angle CRB, computed as the average of all paths' angle CRBs, against the number of paths for various FAA models and their fixed UPA counterparts. Assuming parameters including SNR $=0$ dB, $N_h=N_v=8$, $\lambda=0.03$ meters, $\theta_l\in[\frac{\pi}{3},\frac{2\pi}{3}]$, $\phi_l\in[-\frac{\pi}{3},\frac{\pi}{3}]$, $\alpha_l\sim\mathcal{CN}(0,1)$, and searching for the optimal flexible DoF $\psi$ within $-[\frac{\pi}{2},\frac{\pi}{2}]$, with random parameter generation occurring $1000$ times. As $L$ increases from $1$ to $6$, we observe an increase in angle CRB across all array models with increasing $L$, owing to mutual path effects. Notably, all three proposed FAAs demonstrate significantly lower CRB values compared to the fixed one, irrespective of whether omni-directional or directional antenna patterns are used. This underscores the potential of FAAs for minimizing the CRB in sensing optimization.
 
\vspace{-0.32cm}
	\section{FAA-Enhanced Multi-User Sum-Rate}\label{MUSR}

In this section, the sum-rate metric is employed to further assess the potential of FAAs and evaluate how the flexible DoF introduced by array rotating, bending, or folding can enhance performance.
\vspace{-0.4cm}
\subsection{Single-Sector Sum-Rate}\label{SS}

Firstly, we consider the widely studied single-sector case, where one FAA simultaneously serves $K$ users with precoding based on perfect knowledge of the channel information. 

We adopt the ZF precoding, aiming to derive the precoding matrix $\mathbf{F}\in\mathbb{C}^{N\times K}$ with dual objectives of minimizing total transmit power and effectively eliminating inter-user interference. By stacking all users' channel vectors, we obtain 
\begin{equation}
	\mathbf{H}(\psi)=\left[\overline{\mathbf{h}}_1 (\psi), \overline{\mathbf{h}}_2 (\psi),\cdots,\overline{\mathbf{h}}_K (\psi)\right]\in\mathbb{C}^{N\times K}.
\end{equation} 

The ZF precoder with respect to $\psi$ is given by
\begin{equation}
	\mathbf{F}(\psi)=\mathbf{H}(\psi)\left( \mathbf{H}^H(\psi)\mathbf{H}(\psi)\right)^{-1}.
\end{equation}  
 
Under the total power constraint and with power allocation, the achievable sum-rate can be maximized by:
\begin{equation}\label{mx}
	\begin{aligned}
		\underset{\psi,\{p_k\}_{k=1}^K}{{\rm arg \ max}}& \ \sum_{k=1}^{K} \log_2(1+\gamma_k(\psi) p_k) \\
		{\rm s.t.} \ & \sum_{k=1}^K p_k\leq P, \\
		&\ p_k>0, \forall k,
	\end{aligned}
\end{equation}
where $\gamma_k(\psi)\triangleq \left\Vert [\mathbf{F}(\psi)]_{:,k}\right\Vert_2^{-2}=\frac{1}{\left[\left(\mathbf{H}^H(\psi)\mathbf{H}(\psi)\right)^{-1}\right]_{k,k}}$, $P$ is the total transmit power, and $p_k$ denotes the power allocated for the $k$-th user.

 Notably, $\psi$, which influences $\gamma_k(\psi)$, affects user channel gains, which are crucial for maximizing the objective function. Applying the equal power allocation for simplicity, the maximization of sum-rate with respect to $\psi$ is formulated by
\begin{equation}\label{mx2}
	\begin{aligned}
		\underset{\psi}{{\rm arg \ max}}& \ \sum_{k=1}^{K} \log_2\left(1+\frac{P}{K\sigma_n^2\left[\left(\mathbf{H}^H(\psi)\mathbf{H}(\psi)\right)^{-1}\right]_{k,k}}\right). 
	\end{aligned}
\end{equation}

 While gradient-based methods can solve this problem, they often lead to local optimization. To address this issue, we adopt BO, as described in Section \ref{BO_FAA}.

\subsection{Multi-Sector Sum-Rate}\label{MS}
For more practical systems, a multi-sector BS is typically employed for $360^\circ$ coverage. Utilizing the channel model described in Section \ref{CM}, the multi-sector channel $\overline{\mathbf{H}}(\bm{\psi})\in\mathbb{C}^{3N\times 3K}$, related to the flexible DoF $\psi$, can be re-written as:
\begin{equation}\label{HS}
	\overline{\mathbf{H}}(\bm{\psi})=
	\begin{bmatrix}
		\mathbf{H}_{1,1}(\psi_1)&	\mathbf{H}_{1,2}(\psi_1)&	\mathbf{H}_{1,3}(\psi_1) \\
			\mathbf{H}_{2,1}(\psi_2)&	\mathbf{H}_{2,2}(\psi_2)&	\mathbf{H}_{2,3}(\psi_2)\\
				\mathbf{H}_{3,1}(\psi_3)&	\mathbf{H}_{3,2}(\psi_3)&	\mathbf{H}_{3,3}(\psi_3)
	\end{bmatrix},
\end{equation} 
where $\bm{\psi}\triangleq [\psi_1,\psi_2,\psi_3]$ contains three flexible DoFs, with each corresponding to a FAA. 

For the multi-sector BS, we consider three cooperation strategies for striking flexible tradeoffs between the complexity and achievable performance.

\subsubsection{Separate Flexible Precoding (SFP)}
The sector-separate SINR in Eqn. (\ref{SINRS0}) can be re-written as ${\rm SINR}_{m_k}(\psi_m)$ by considering the impact of the flexible DoFs on the channel, i.e., ${\mathbf{h}}_{m,m_k}\rightarrow\overline{\mathbf{h}}_{m,m_k}(\psi_m)$.

Using the ZF prcoding for each sector individually, akin to Eqn. (\ref{mx2}), yields $\mathbf{H}_{m,m}^H(\psi_m)\mathbf{F}_{m}=\mathbf{I}_K$. However, this results in $\mathbf{H}_{m,m^\prime}^H(\psi
_m)\mathbf{F}_{m}\neq\mathbf{I}_K, m\neq m^\prime$. 
 In this sense, the $k$-user in the $m$-th sector is only interfered by other sectors and noise. Hence, ${\rm SINR}_{m_k}(\psi_m)$ under ZF with equal power allocation can be simplified into 
\begin{equation}\label{SI2}
	\begin{aligned}{\rm SINR}_{m_k}(\psi_m)= &  \frac{P\left[\left(\mathbf{H}_{m,m}^H(\psi_m)\mathbf{H}_{m,m}(\psi_m)\right)^{-1}\right]_{k,k}^{-1}}{3K\sum_{m^\prime\neq m}^3 \sum_{i}^{K}\left\vert\overline{\mathbf{h}}_{m^\prime,m_k}^H\mathbf{f}_{m^\prime_i}\right\vert^2+\sigma_n^2} \\
		=&\frac{P}{3K\overline{\sigma}_n^2\left[\left(\mathbf{H}_{m,m}^H(\psi_m)\mathbf{H}_{m,m}(\psi_m)\right)^{-1}\right]_{k,k}},
	\end{aligned} 
\end{equation}
where $\overline{\sigma}_n^2\triangleq\sum_{m^\prime\neq m}^3 \sum_{i}^{K}\left\vert\overline{\mathbf{h}}_{m^\prime,m_k}^H\mathbf{f}_{m^\prime_i}\right\vert^2+\sigma_n^2$.

Hence, for the $m$-th sector, the optimal $\psi_m^\star$ is solved by
\begin{equation}\label{mx22}
\psi^\star_m=	\underset{\psi_m}{\rm arg\ max} \ \sum_{k=1}^{K} \log_2\left( 1+ {\rm SINR}_{m_k}(\psi_m) \right).
\end{equation}

Finally, the total multi-sector sum-rate is calculated by $ \sum_{m=1}^M\sum_{k=1}^{K} \log_2\left( 1+ {\rm SINR}_{m_k}(\psi^\star_m) \right)$.
\subsubsection{Joint Flexible Precoding (JFP)}
As all sector antennas are assumed to be connected by a CPU, facilitating joint processing, we can execute sector-joint processing on all antennas for all users, which involves precoding on the entire channel as depicted in Eqn. (\ref{HS}). Consequently, the sector-joint SINR facilitated by the three-sector FAAs, as expressed in Eqn. (\ref{SINRJ0}), can be reformulated in terms of $\bm\psi$:
\begin{equation} 
	\begin{aligned}
		{\rm SINR}_{m_k}(\bm{\psi})=\frac{\sum_{m=1}^3\left\vert\overline{\mathbf{h}}_{m,m_k}^H\mathbf{f}_{m_k}\right\vert^2}{\sum_{m^\prime\neq m}^3 \sum_{i\neq k}^{K}\left\vert\overline{\mathbf{h}}_{m^\prime,m_k}^H\mathbf{f}_{m^\prime_i}\right\vert^2+\sigma_n^2}.
	\end{aligned}
\end{equation}

Applying ZF precoding with equal power allocation, similar to Eqn. (\ref{mx2})
, the maximization of sum-rate with respect to $\bm{\psi}$ is formulated by
\begin{equation}\label{mx3}
	\begin{aligned}
		\underset{\bm{\psi}}{{\rm arg \ max}}& \ \sum_{k=1}^{3K} \log_2\left(1+\frac{P}{3K\sigma_n^2\left[\left(\overline{\mathbf{H}}^H(\bm{\psi})\overline{\mathbf{H}}(\bm{\psi})\right)^{-1}\right]_{k,k}}\right).
	\end{aligned}
\end{equation}

Compared to Eqns. (\ref{mx2}) and (\ref{mx22}), the sector-joint optimization problem is found to encompass all FAAs' DoFs, thus involving a joint optimization of FAA shapes. 

\subsubsection{Semi-Joint Flexible Precoding (SJFP)}
It is worth noting that joint processing necessitates the use of all users' channels for ZF precoding, potentially leading to high complexity due to the inversion operation. On the other hand, separate processing involves ZF precoding only over users within each sector, but this might result in high interference for edge users if the element pattern is not ideal.

Therefore, to strike a balance between performance and complexity, we introduce a semi-joint method that combines separate precoding with joint array shape optimization. This approach effectively suppresses interference from other sectors to some context.
To elaborate, we initially execute ZF precoding for users within each sector, which mirrors the sector-separate processing. According to Eqn. (\ref{SI2}), we can obtain
\begin{equation}\label{SI4}
	{\rm SINR}_{m_k}(\bm{\psi})=   \frac{P\left[\left(\mathbf{H}_{m,m}^H(\psi_m)\mathbf{H}_{m,m}(\psi_m)\right)^{-1}\right]_{k,k}^{-1}}{3K\sum_{m^\prime\neq m}^3 \sum_{i}^{K}\left\vert\overline{\mathbf{h}}_{m^\prime,m_k}^H(\psi_{m^\prime})\mathbf{f}_{m^\prime_i}\right\vert^2+\sigma_n^2}.
\end{equation}

Comparing with Eqn. (\ref{SI2}), we can find that the  denominator part of Eqn. (\ref{SI4}) is also correlated with other sector FAAs' $\psi$. This identifies that we can optimize $\psi_{m^\prime}$ to suppress the interference from the $m$-th sector. Then, the multi-sector sum-rate optimization problem is expressed by
\begin{equation}
	\underset{\bm{\psi}}{\rm arg\ max} \ \sum_{m=1}^M \sum_{k=1}^{K} \log_2\left( 1+ {\rm SINR}_{m_k}(\bm{\psi}) \right).
\end{equation}

In summary, this semi-joint processing approach necessitates sector-dimensional ZF precoding, as opposed to precoding across all sector antennas in the joint processing method. Furthermore, it efficiently mitigates high inter-sector interference by capitalizing on the flexible DoF.
\vspace{-0.2cm}
\section{BO for FAA Optimization}\label{BO_FAA}

Thanks to the closed-form expression of ZF precoder, the equal-power allocation sum-rate in Sections \ref{SS} and \ref{MS} depends solely on the flexible DoFs, such as rotating angles or bending angles. However, obtaining a closed-form solution for $\bm{\psi}$ is challenging due to its complex expression.
Instead, efficient solutions can be obtained using search methods. One commonly used approach is gradient descent/ascent, which relies on the gradient and step size to iteratively search for an optimal solution. However, this method is sensitive to the initial point and may converge to a local optimum. Additionally, calculating the gradient in some complex cases may lead to high computational complexity. Another method is global optimization, which can achieve high performance. In this study, BO is employed to solve optimal rotating or bending angles in conjunction with ZF precoding.

Although there exist several studies on BO for diverse applications \cite{BO1,BO2,BO3}, we focus on BO in this context, integrating sum-rate and our insights.
To prevent confusion, the BO approach discussed in this study relies on Gaussian processes (GPs) for surrogate modeling and expected improvement (EI) for acquisition functions. To enhance clarity, we consider the single-sector sum-rate for formula derivation, corresponding to the objective function $f$ defined in Eqn. (\ref{mx2}). This framework can be readily extended to multi-sector sum-rate scenarios.
 \begin{figure*}
	\centering
	\subfigure[Iteration 1]{
		\includegraphics[width=2.16in]{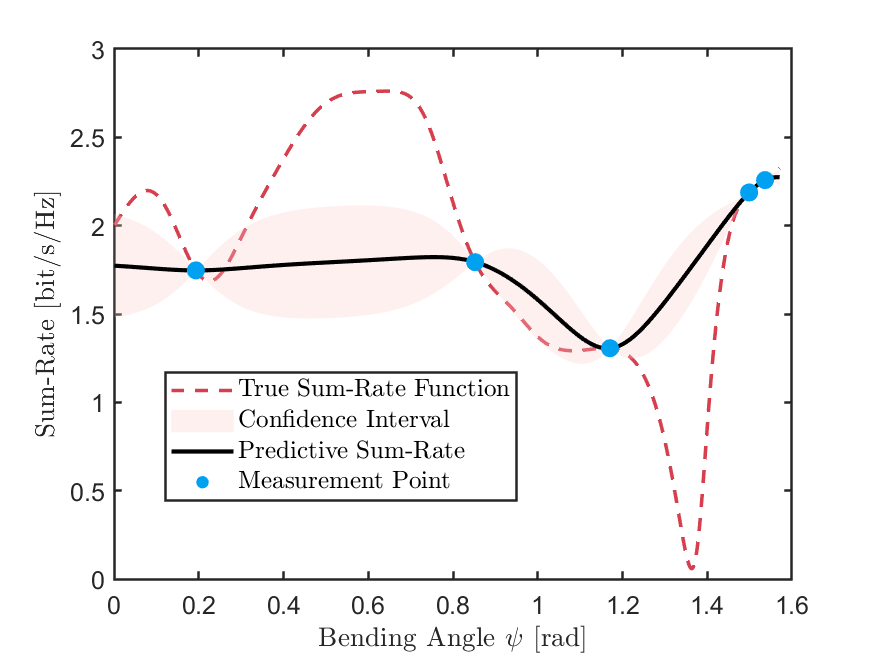}
	}
	\quad    
	\subfigure[Iteration 2]{
		\includegraphics[width=2.16in]{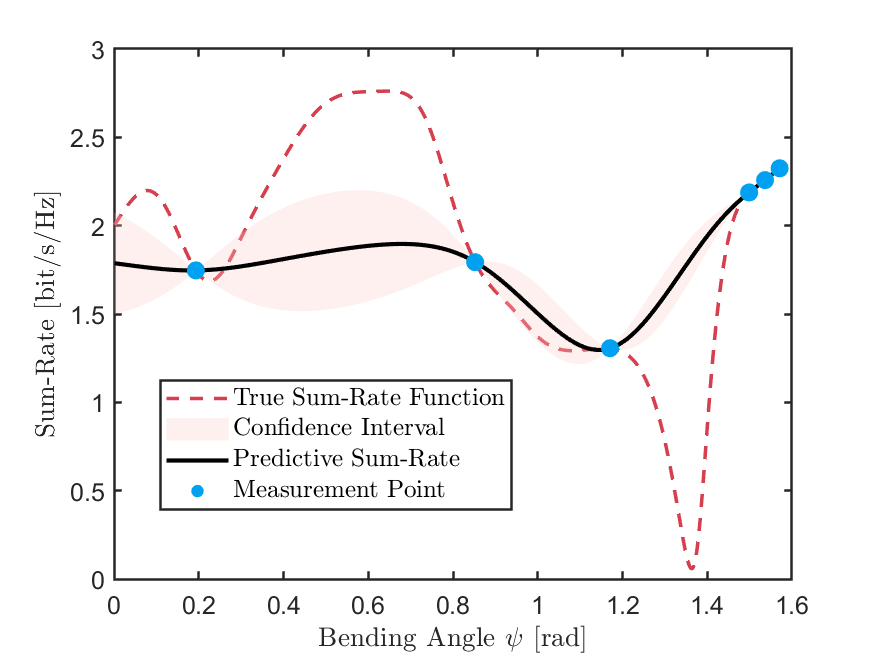}
	}
	\subfigure[Iteration 3]{
		\includegraphics[width=2.16in]{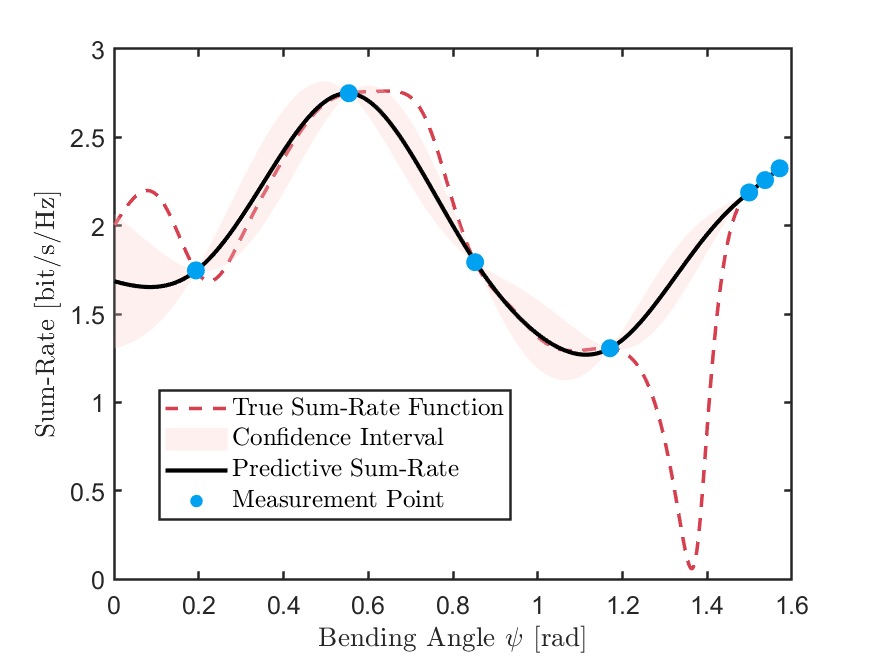}
	}
	\caption{The predictive trend of three BO iterations with four inital measurement points.}
	\label{BOI}
\end{figure*}
\subsection{GP Regression}

A random process consists of an infinite number of random variables. GP assumes that all these random variables follow Gaussian distributions, each potentially having different means and correlated variances. Notably, the correlated variances aid in describing the underlying relationship among the function values of closely related variables. We define a GP with respect to the continuous Gaussian random variable $\psi$, characterized by a mean function $\mu(\psi)$ and a covariance function $\mathcal{K}(\psi,\psi^\prime)$. The covariance function can take various forms, such as the  squared exponential function $\mathcal{K}\left(\psi,\psi^\prime\right)=\eta_0\exp \left(-\frac{\eta_1}{2}|\psi-\psi^\prime|^2\right)$, where $\eta_0$ and $\eta_1$ denote hyperparameters that control the kernel.

We first discuss the GP regression for Eqn. (\ref{mx2}) which uses a GP model to fit a function. Let us consider the true function $f(\psi)$ with respect to $\psi$:
 \begin{equation}\label{fp}
 	\begin{aligned}
 	f(\psi) \triangleq \sum_{k=1}^{K} \log_2\left(1+\frac{P}{K\left[\left(\mathbf{H}^H(\psi)\mathbf{H}(\psi)\right)^{-1}\right]_{k,k}}\right)
 	\end{aligned}.
 \end{equation}
Consider $T$ sampling points $\widetilde{\bm{\psi}}\triangleq[\psi^{(1)},\psi^{(2)},\cdots,\psi^{(T)}]$ for the objective function $f$, with corresponding measurement values $f\left(\widetilde{\bm{\psi}}\right)\triangleq\left[f(\psi^{(1)}),f(\psi^{(2)}),\cdots,f(\psi^{(T)})\right]$. The covariance of $\widetilde{\bm{\psi}}$ is given by
\begin{equation}
	\mathbf{K}\left(\widetilde{\bm{\psi}},\widetilde{\bm{\psi}}\right)\triangleq	\begin{bmatrix}
	\mathcal{K}\left(\psi^{(1)},\psi^{(1)}\right) &\cdots &	\mathcal{K}\left(\psi^{(1)},\psi^{(T)}\right) \\
		\vdots & \ddots &\vdots \\	\mathcal{K}\left(\psi^{(T)},\psi^{(1)}\right) &\cdots &	\mathcal{K}\left(\psi^{(T)},\psi^{(T)}\right)
	\end{bmatrix}.
\end{equation}

Given the dataset $\bm{\mathcal{D}}\triangleq\left\{\widetilde{\bm{\psi}},f\left(\widetilde{\bm{\psi}}\right)\right\}$, our goal is to predict the objective function. In other words, for any arbitrary $\psi$, we aim to determine its function value $f^\star(\psi)$ that approximates the true value using GP regression. Hence, considering the Gaussian random variable $\psi$ where $\psi\notin\widetilde{\mathbb{\bm{\psi}}}$, the joint distribution of the measured function values $f\left(\widetilde{\bm{\psi}}\right)$ and the predicted value $f^\star(\psi)$ at $\psi$ is given by
\begin{equation}
\begin{bmatrix}
f\left(\widetilde{\bm{\psi}}\right) \\
f^\star(\psi)
\end{bmatrix}  \sim \mathcal{N}\begin{pmatrix}
\overline{{\mu}}\left(\widetilde{\bm{\psi}}\right),\begin{bmatrix}
		\mathbf{K}\left(\widetilde{\bm{\psi}},\widetilde{\bm{\psi}}\right) & 	\mathbf{K}\left(\widetilde{\bm{\psi}},\psi \right) \\ \mathbf{K}^T\left(\widetilde{\bm{\psi}},\psi \right)& \mathbf{K}\left(\psi,\psi \right)
\end{bmatrix}
\end{pmatrix}.
\end{equation}
 
Utilizing the conditional distribution properties of multivariate Gaussian distributions, $p\left(f^\star(\psi) | f\left(\widetilde{\bm{\psi}}\right) \right)$ is given by
\begin{equation}
	\begin{aligned}
		& p\left(f^\star(\psi) | f\left(\widetilde{\bm{\psi}}\right) \right) = \\ & \mathcal{N}\left(   \mathbf{K}^T\left(\widetilde{\bm{\psi}},\psi \right)	\mathbf{K}^{-1}\left(\widetilde{\bm{\psi}},\widetilde{\bm{\psi}}\right)  		\left(f\left(\widetilde{\bm{\psi}}\right)-\overline{{\mu}}\left(\widetilde{\bm{\psi}}\right) \right) + \overline{\mu}(\psi) ,\right. \\ & \left.
\mathbf{K}\left(\psi,\psi \right)- \mathbf{K}^T\left(\widetilde{\bm{\psi}},\psi \right) \mathbf{K}^{-1}\left(\widetilde{\bm{\psi}},\widetilde{\bm{\psi}}\right)  \mathbf{K}\left(\widetilde{\bm{\psi}},\psi \right)	
		 \right) .
	\end{aligned}
\end{equation}

The hyperparameter of the prior distribution can be learned by the maximum a posterior estimate.
 For simplicty,  the zero-mean  $\overline{\mu}(\cdot)=0$ and $\eta_0=\eta_1=1$ can be assumed for the prior distribution.
 \vspace{-0.2cm}
\subsection{EI Acquisition}

According to the analysis before, we could suppose the posterior probablity of the next sampling point follows $p\left(f^\star(\psi^{(T+1)}) | f\left(\widetilde{\bm{\psi}}\right) \right)=\mathcal{N}\left(\mu(\psi^{(T+1)}),\sigma^2(\psi^{(T+1)})\right)$, where
\begin{equation}\label{mu1}
	\mu(\psi^{(T+1)})\triangleq    \mathbf{K}^T\left(\widetilde{\bm{\psi}},\psi^{(T+1)} \right)	\mathbf{K}^{-1}\left(\widetilde{\bm{\psi}},\widetilde{\bm{\psi}}\right)  		 f\left(\widetilde{\bm{\psi}}\right), 
\end{equation}
\begin{equation}\label{sig1}
	\begin{aligned}
		\sigma^2(\psi^{(T+1)})\triangleq & \mathbf{K}\left(\psi^{(T+1)},\psi^{(T+1)} \right)- \mathbf{K}^T\left(\widetilde{\bm{\psi}},\psi^{(T+1)} \right) \\ & 
		\ \mathbf{K}^{-1}\left(\widetilde{\bm{\psi}},\widetilde{\bm{\psi}}\right)  \mathbf{K}\left(\widetilde{\bm{\psi}},\psi^{(T+1)} \right).
	\end{aligned}	
\end{equation}

With the EI strategy \cite{BO3}, we have

\begin{equation}\label{EIM}
\psi^{(T+1)} = \underset{\psi^{(T+1)}}{\rm arg \ max}
\ \mathbb{E} \left\{ \mathcal{I}\left(\psi^{(T+1)} , f^{(T),\star}
\right) | \bm{\mathcal{D}}_T
\right\},
\end{equation}
where $\mathcal{I}\left(\psi^{(T+1)} , f^{(T),\star}
\right)\triangleq {\rm max}\left\{0, f\left(\psi^{(T+1)}
\right)-f^{(T),\star}
\right\}$ denotes the improvement function, $f^{(T),\star}$ represents the maximum function value of the $T$ measurements, i.e., $f^{(T),\star}\triangleq {\rm max}\left\{f\left(\widetilde{\bm{\psi}}\right)\right\}$.

Since $\psi^{(T+1)}$ is a Gaussian random variable, the EI function in Eqn. (\ref{EIM}) is written as
\begin{equation}
	\begin{aligned}
	&	\mathbb{E} \left\{ \mathcal{I}\left(\psi^{(T+1)} , f^{(T),\star}
		\right) | \bm{\mathcal{D}}_T
		\right\}=\\
		& \frac{1}{\sigma \left(\psi^{(T+1)}\right) \sqrt{2\pi}} \int_{-\infty}^{\infty}   {\rm max}\left\{0, b
	 -f^{(T),\star}	\right\} e^{-\frac{\left(b-\mu\left(\psi^{(T+1)}\right)\right)^2}{2\sigma^2\left(\psi^{(T+1)}\right)}}
	{\rm d} b \\
	&= \frac{1}{\sigma \left(\psi^{(T+1)}\right) \sqrt{2\pi}} \int_{f^{(T),\star}}^{\infty} \left(b
	-f^{(T),\star}	\right) e^{-\frac{\left(b-\mu\left(\psi^{(T+1)}\right)\right)^2}{2\sigma^2\left(\psi^{(T+1)}\right)}}
	{\rm d} b \\
	&= \sigma \left(\psi^{(T+1)}\right)  \varphi \left(\frac{\mu \left(\psi^{(T+1)}\right)-f^{(T),\star}}{\sigma \left(\psi^{(T+1)}\right)}\right)\\& \ \ \ \ + \left(\mu \left(\psi^{(T+1)}\right)-f^{(T),\star}\right)\Phi \left(\frac{\mu \left(\psi^{(T+1)}\right)-f^{(T),\star}}{\sigma \left(\psi^{(T+1)}\right)}\right),
	\end{aligned}
\end{equation}
where $\varphi(a)$ and $\Phi(a)$ denote the probability density function and cumulative distribution function, respectively, given by
\begin{equation}
	\begin{aligned}
	 \varphi(a)=\frac{1}{\sqrt{2 \pi}} e^{-\frac{a^2}{2}} ,
	\end{aligned}
\end{equation}
\begin{equation} \Phi(a)=\frac{1}{\sqrt{2 \pi}} \int_{-\infty}^a e^{-\frac{t^2}{2}} {\rm d} t.
\end{equation}

Therefore, the EI acquisition function $\mathcal{L}(\cdot)$ with respect to $\psi$ can be summarized as
\begin{equation}\label{EI}
\mathcal{L}(\mathbf{\psi})= \begin{cases} 
	\left(\mu(\psi)-f^\star \right) \Phi(\Psi)+\sigma(\psi) \varphi(\Psi), & \sigma(\psi)>0 \\
	0, & \sigma(\psi)=0  
\end{cases}
\end{equation}
where $\Psi=\frac{\mu(\psi)-f\left(\mathbf{\psi}^{\star}\right)}{\sigma(\psi)}$.
\vspace{-0.2cm}
\subsection{Incremental Search}
 
 Building upon GP regression and the EI acquisition function, we can construct an incremental search strategy to find the global optimal solution for $\psi$. In the case of multi-sector scenarios, the variable of interest becomes $\bm{\psi}\triangleq{\psi_1,\psi_2,\psi_3}$.
 
 Initially, we randomly select $T$ points $\widetilde{\bm{\psi}}$ to measure the function $f\left(\widetilde{\bm{\psi}}\right)$, thereby generating the dataset $\bm{\mathcal{D}}_T$. Subsequently, by combining Eqs. (\ref{mu1}), (\ref{sig1}), and (\ref{EIM}), we identify the $(T+1)$ optimal measurements. We then update the dataset to $\bm{\mathcal{D}}_{T+1}$ and proceed with the next point evaluation following the same procedure.
 
 For a more comprehensive understanding of the BO process, we utilize graphical representations of GP regression outcomes, as depicted in Fig. \ref{BOI}. Within this framework, we evaluate the single-sector sum-rate using Eqn. (\ref{fp}) with BO, contrasting it with the true sum-rate function obtained from exhaustive search, visually depicted by the red dashed line. While exhaustive search over $\psi\in[0,\frac{\pi}{2}]$ for Eqn. (\ref{fp}) yields the ideal function value, this approach is impractical in real-world systems, particularly for multi-sector scenarios. For clarity, only the bendable FAA is considered.

Initiating BO with four random measurement points, we subsequently utilize the EI acquisition function to identify the next optimal measurement point. This process guides us through Gaussian regression steps, as demonstrated in Fig. \ref{BOI}(a). Notably, this iteration of BO identifies a local near-optimal point along the red dashed line.
In the second iteration, as depicted in Fig. \ref{BOI}(b), a neighboring point around $\psi=1.57$ is measured, emphasizing ``Exploitation", aimed at approaching the local optimum within the measured region while the confidence interval between measurement points diminishes.
Transitioning to Fig. \ref{BOI}(c), the EI strategy prompts a measurement around $\psi=0.55$, signifying ``Exploration", which involves seeking unexplored areas within the search space that may harbor superior solutions. Despite not precisely predicting the true function, BO manages to find the global optimum in the current iteration, as evidenced by the comparison between the black line and the red dashed line.

Balancing exploration and exploitation stands as a critical aspect of BO. Overemphasis on exploration may lead to scattered searches with insufficient informative content, whereas excessive exploitation might result in entrapment within local optima, overlooking the global optimum. Hence, BO algorithms dynamically adjust the exploration-exploitation trade-off to enhance optimization performance.

\vspace{-0.2cm}
\section{Simulation Results}\label{SIMR}
This section conducts several numerical simulation to evaluate FAAs' improment in terms of sum-rate.  The simulations are set in a system operating at a central frequency of $10$ GHz. The BS is equipped with $N = 8 \times 2$ antennas for each sector and serves $K=N$ users. Path gain $\beta_{k,l}$, for all $k,l$, is modeled following a complex Gaussian distribution $\mathcal{CN}(0,1)$. 
Users and scatterers are uniformly distributed across a $100$-meter circle, conforming to the far-field assumption. All users have elevation angles $\theta_{k,l}\in[\frac{\pi}{3},\frac{2\pi}{3}]$. Users within the first sector have azimuth angles uniformly distributed in $[-\frac{\pi}{3},\frac{\pi}{3}]$, those within the second sector have azimuth angles uniformly distributed in $[\frac{\pi}{3},\pi]$, and users within the third sector have azimuth angles uniformly distributed in $[\pi,\frac{5\pi}{3}]$.
 The number of channel paths, $L$, is fixed at $5$. The noise power $\sigma^2$ is set to $1$, and the SNR is defined as $P/\sigma^2$.

The simulation methods involve:
\begin{itemize}
	\item \textbf{SZF-UPA, JZF-UPA, and SJZF-UPA:} 
	They utilize fixed UPAs, corresponding to the three multi-sector processing approaches outlined in Section \ref{MS}: SFP, JFP, and SJFP, all without the flexibility of DoF.
	\item \textbf{SFP-Rotating, JFP-Rotating, and SJFP-Rotating:} 
	They utilize the rotatable FAA model from Section \ref{FRM} for the three multi-sector processing methods outlined in Section \ref{MS}, solved by BO with variables. $\psi_1,\psi_2,\psi_3\in[-\frac{\pi}{4},\frac{\pi}{4}]$.
		\item \textbf{SFP-Bending, JFP-Bending, and SJFP-Bending:} 
		They utilize the bendable FAA model from Section \ref{FBM} for the three multi-sector processing methods outlined in Section \ref{MS}, solved by BO with variables. $\psi_1,\psi_2,\psi_3\in[-\frac{\pi}{2},\frac{\pi}{2}]$.
			\item \textbf{SFP-Folding, JFP-Folding, and SJFP-Folding:} 
			They utilize the foldable FAA model from Section \ref{FFM} for the three multi-sector processing methods outlined in Section \ref{MS}, solved by BO with variables. $\psi_1,\psi_2,\psi_3\in[-\frac{\pi}{4},\frac{\pi}{4}]$.
\end{itemize}

\begin{figure}
	\centering
	\subfigure[Omni-directional pattern.]{
		\includegraphics[width=2.8in]{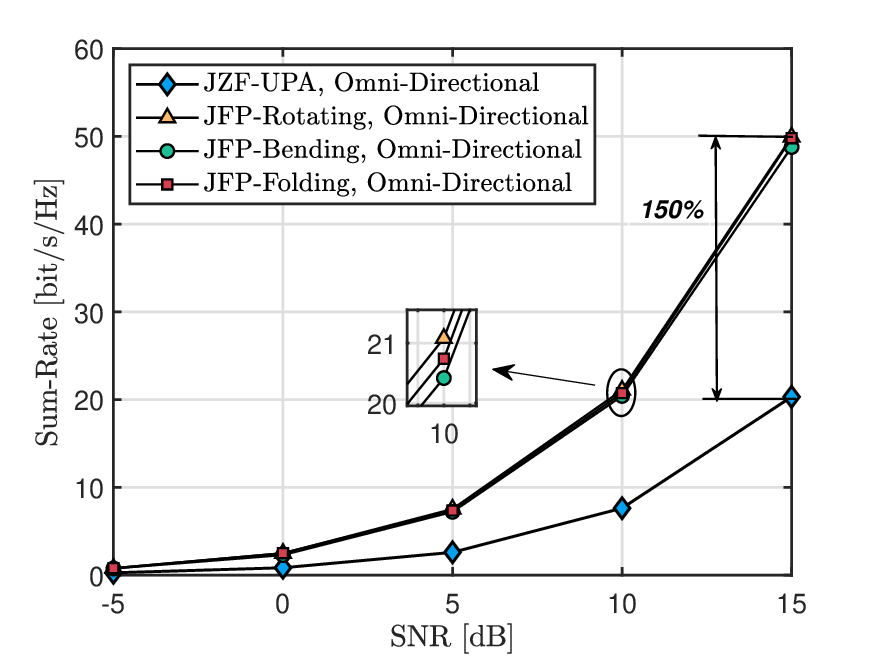}
	}
	\\    
	\subfigure[Directional pattern with $\kappa=1$ and $\kappa=2$.]{
		\includegraphics[width=2.8in]{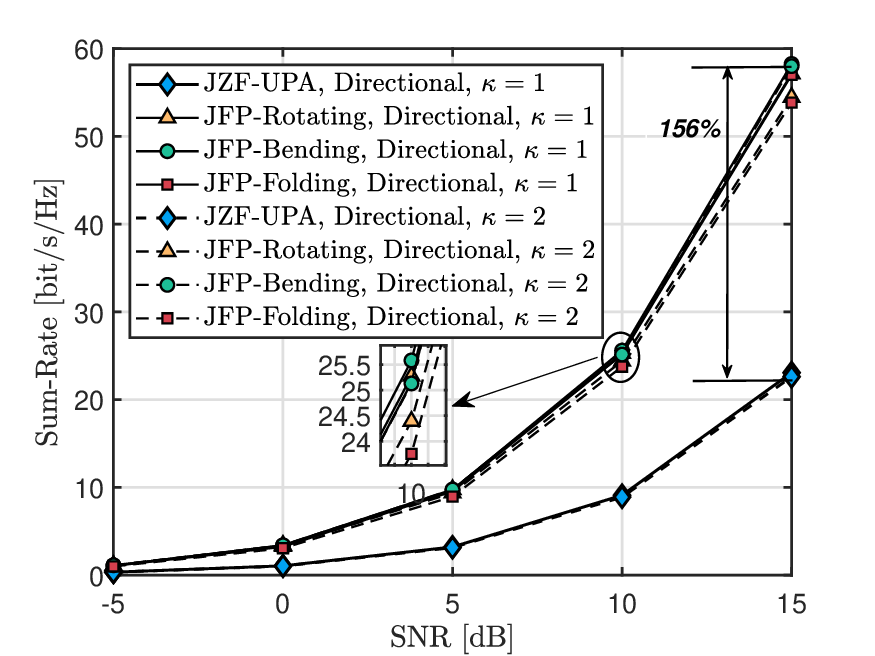}
	} 
	\caption{JFP-based sum-rate of FAAs and UPAs.}
	\label{JFP_plot}
\end{figure}

Fig. \ref{JFP_plot} evaluates the joint processing approach for three-sector sum-rate, considering both omni-directional and directional patterns, with SNR ranging from 0 to 15 dB. From Fig. \ref{JFP_plot}(a), where omni-directional patterns are used for all models, it can be observed that the three FAA models exhibit comparable performance, with the rotatable FAA slightly outperforming the others. Notably, at SNR $=15$ dB their sum-rate performance is approximately $250\%$ of that of JZF-UPA without flexible DoFs, or about $150\%$ higher than JZF-UPA. This notable improvement is primarily attributed to the antenna position mapping, which varies with flexible DoFs and aids the three-sector BS in adjusting antenna positions to improve channel conditions, thereby enabling better user interference cancellation.

When considering the directional pattern discussed in Section \ref{ARP} with $\kappa=1$ and $\kappa=2$, Fig. \ref{JFP_plot}(b) shows that all schemes exhibit performance improvements compared to those in Fig. \ref{JFP_plot}(a), indicating the usefulness of the directional pattern in this scenario, especially for FAAs. For instance, the bendable FAA experiences an approximately $18\%$ performance improvement, enabling it to outperform other schemes. Moreover, in this case, the best FAA performance achieves about a $156\%$ improvement over the fixed UPA. This is attributed to the importance of antenna orientation provided by the directional pattern, which serves as a DoF for performance optimization. Notably, $\kappa=2$, which offers a more directional beam than $\kappa=1$, exhibits slight performance degradation in all methods with joint processing, suggesting that stronger directivity in the directional pattern does not necessarily lead to better sum-rate performance in this scenario.

\begin{figure}
	\centering
	\subfigure[Omni-directional pattern.]{
		\includegraphics[width=2.8in]{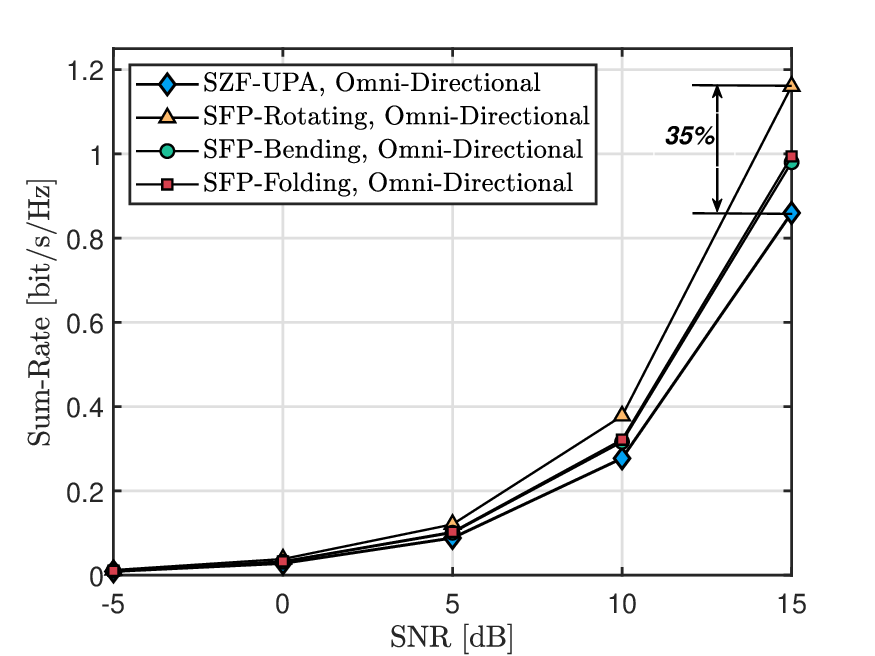}
	}
	\\    
	\subfigure[Directional pattern with $\kappa=1$ and $\kappa=2$.]{
		\includegraphics[width=2.8in]{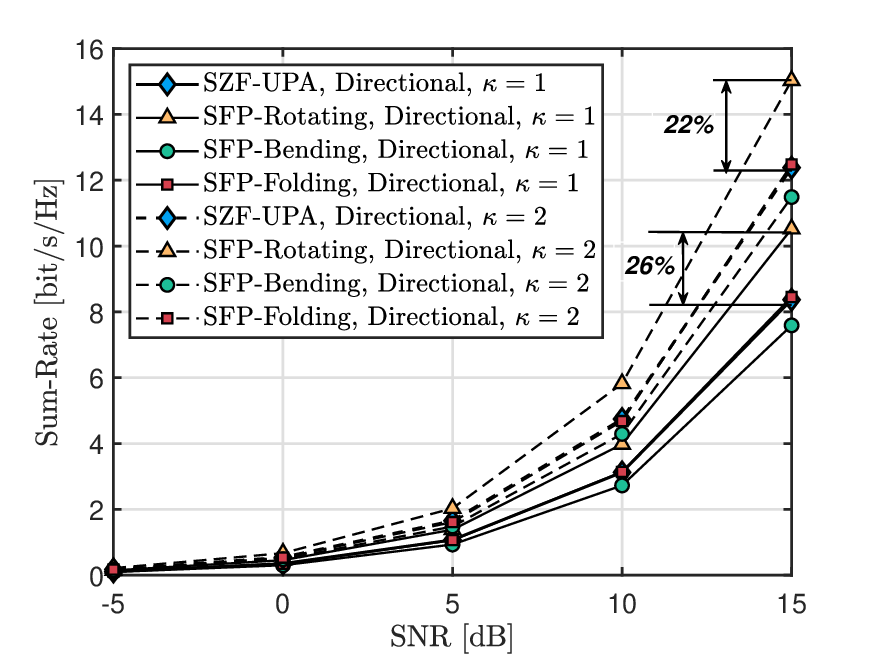}
	} 
	\caption{SFP-based sum-rate of FAAs and UPAs.}
	\label{SFP_plot}
\end{figure}

As analyzed in Section \ref{MS}, separate processing for the sectorize BS has low complexity since each FAA only serves users within its sector. However, this causes high interference from other sectors. Fig. \ref{SFP_plot}(a) clarifies this by setting the omni-directional pattern for all antennas. We can find that it shows a very poor performance since for arbitrary one sector's users, other two sector's precoding will significantly impact them due to the omni-directional radiation. On the other hand, when adopting directional patterns in Fig. \ref{SFP_plot}(b), the sum-rate performance increases significantly compared to Fig. \ref{SFP_plot}(a). This is because the directional pattern could enable each sector FAA focuses the antenna power toward users within this sector and reduces the engergy radiated towards other sectors, thereby decreasing the inter-user interference among sectors. In this case, we can see that the directivity of the pattern plays an important role as $\kappa=2$ with the dashed line outperforms $\kappa=1$ with the solid line in different methods. This is due to the fact that a larger value of $\kappa$ can better assist the array of a certain sector to radiate only in that sector, which is desirable in the separate processing mode. In particular, rotatable FAA in this scenario can also make a good performance improvement. However, bendable and foldable has no obvious improvement for this case, as each sector's FAA considers only one DoF without strong controlling capability. Bending or folding could lead to certain antenna orientations optimized for the current sector's sum-rate, inadvertently directing radiation towards other sectors and causing interference to users within those sectors. Therefore, performance improvement is not guaranteed when calculating all sectors' sum-rate.

\begin{figure}
	\centering
	\subfigure[Omni-directional pattern.]{
		\includegraphics[width=2.8in]{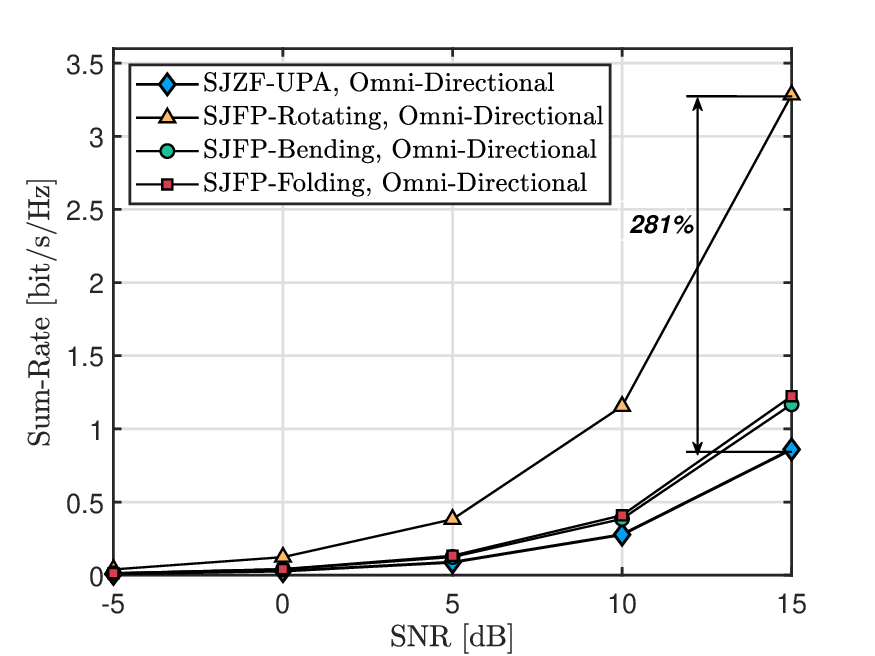}
	}
	\\    
	\subfigure[Directional pattern with $\kappa=1$ and $\kappa=2$.]{
		\includegraphics[width=2.8in]{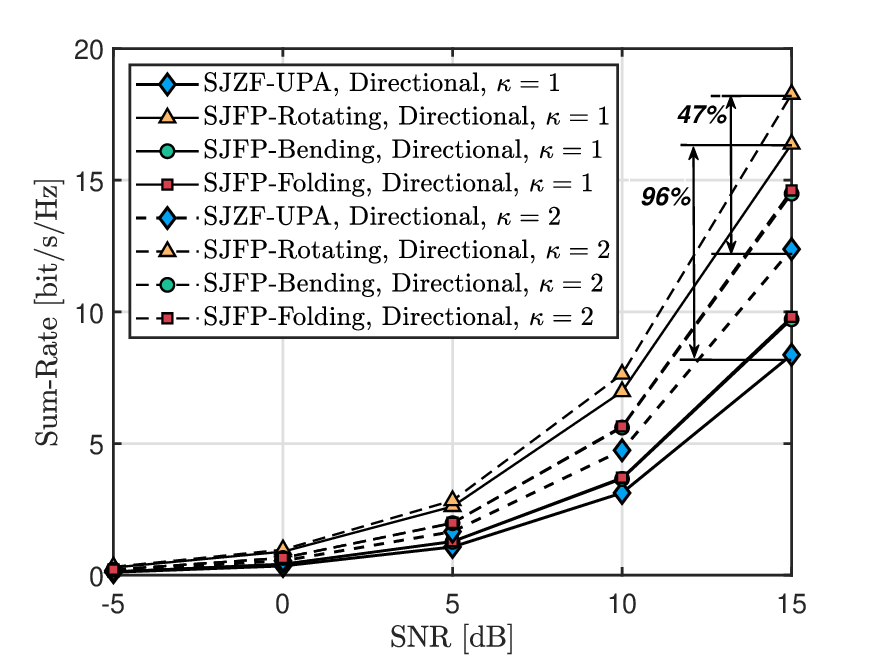}
	} 
	\caption{SJFP-based sum-rate of FAAs and UPAs.}
	\label{SJFP_plot}
\end{figure}

As a modified strategy, SJFP strikes a balance between SFP and JFP, as evaluated in Fig. \ref{SJFP_plot}. Similarly, both omni-directional and directional patterns are considered. In Fig. \ref{SJFP_plot}(a), despite its relatively modest performance, SJFP shows some improvement compared to the separate processing strategy depicted in Fig. \ref{SFP_plot}(a), particularly noticeable for the rotatable FAA, which achieves a $181\%$ increase in sum-rate compared to the fixed UPA. It is worth noting that this semi-joint sector processing is exclusively applicable to FAAs and has no impact on the fixed case. Comparing Fig. \ref{SJFP_plot}(a) with Fig. \ref{SJFP_plot}(b), the latter exhibits significantly improved performance, attributed to the directional pattern. Furthermore, as the directivity strength ($\kappa$) increases from $1$ to $2$, the performance enhancement becomes more pronounced. Specifically, in Fig. \ref{SJFP_plot}(b), the rotatable FAA achieves performance gains of $47\%$ and $96\%$ compared to the fixed case with $\kappa=1$ and $\kappa=2$, respectively, highlighting the potential of rotatable FAAs in this semi-joint processing scenario. This implies the unique capability of the rotatable FAA to enhance the user sum-rate within its own sector while suppressing interference on users from other sectors.
\section{Conclusions}\label{Con}

This paper explores the capabilities of FAAs featuring horizontally rotatable, bendable, and foldable shapes, providing dynamic control over antenna positions and orientations. Initially, the positions and orientations of array elements influenced by the flexible DoF caused by rotation, bending, and folding are mathematically modeled. This process redefines the multi-user channel with respect to the flexible DoF, introducing an additional DoF for wireless communications. The impact of the flexible DoF on performance optimization, considering both omni-directional and directional patterns, is further investigated. This involves reassessing the increase in channel power and decrease in channel angle CRB by optimizing the flexible DoF for the three FAA models.
Moreover, the FAA's potential to enhance multi-sector sum-rate is evaluated. To this end, this work proposes three strategies: SFP, JFP, and SJFP, all of which fully utilize the FAAs' DoFs. In all three cases, the directional pattern demonstrates superior performance compared to the omni-directional pattern, particularly for JFP and SJFP. Notably, the three FAA models exhibit significant sum-rate enhancements in certain scenarios. For instance, the bendable FAA achieves a $156\%$ gain compared to the UPA in the case of JFP with the directional pattern. Furthermore, the rotatable FAA exhibits notably superior performance in SFP and SJFP cases with omni-directional patterns, with respective $35\%$ and $281\%$.

\bibliographystyle{IEEEtran}
\bibliography{reference.bib}

\vspace{12pt}

\end{document}